\documentclass[twocolumn,useAMS,usenatbib]{mn2e}
\usepackage{graphicx,amsmath}
\usepackage{bm}

\newcommand{\rmd}{{\rm d}}
\newcommand{\rme}{{\rm e}}
\newcommand{\rmi}{{\rm i}}

\newcommand{\rpi}{\Pi}

\newcommand{\beq}{\begin{eqnarray}}
\newcommand{\eeq}{\end{eqnarray}}
\newcommand{\beqa}{\begin{eqnarray}}
\newcommand{\eeqa}{\end{eqnarray}}
\newcommand{\hkpc}{$h^{-1}\,$kpc}
\newcommand{\hmpc}{$h^{-1}\,$Mpc}

\newcommand{\BCG}{BG}
\newcommand{\nonBCG}{non-BG}

\title[Intrinsic alignments]
{Intrinsic galaxy alignments from the 2SLAQ and SDSS surveys: luminosity and redshift scalings and 
implications for weak lensing surveys}

\author[Hirata et al.]
 {Christopher M. Hirata$^1$\thanks{Electronic address:
    {\tt chirata@sns.ias.edu}},
  Rachel Mandelbaum$^1$\thanks{Hubble Fellow.},
  Mustapha Ishak$^2$,
\newauthor
  Uro\v s Seljak$^{3,4,5}$,
  Robert Nichol$^6$,
  Kevin A. Pimbblet$^7$,
  Nicholas P. Ross$^8$, and
\newauthor
  David Wake$^8$
\\$^1$Institute for Advanced Study, Einstein Drive, Princeton, NJ 08540, USA
\\$^2$Department of Physics, The University of Texas at Dallas, Richardson, TX 75083, USA
\\$^3$Institute for Theoretical Physics, University of Zurich, 8057 Zurich, Switzerland
\\$^4$International Centre for Theoretical Physics, Strada Costiera 11, 34014 Trieste, Italy
\\$^5$Department of Physics, Princeton University, Princeton, NJ 08544, USA
\\$^6$Institute of Cosmology and Gravitation, Mercantile House, University of Portsmouth, Portsmouth, PO1 2EG, UK
\\$^7$Department of Physics, University of Queensland, Brisbane, QLD 4072, Australia
\\$^8$Physics Department, University of Durham, South Road, Durham, DH1 3LE, UK
}

\date{\today}

\begin{document}
\maketitle

\begin{abstract}
Correlations between intrinsic shear and the density field on large scales, a potentially important contaminant for cosmic shear surveys, have been 
robustly detected at low redshifts with bright galaxies in SDSS data. Here we present a more detailed characterization of this effect, which can cause 
anti-correlations between gravitational lensing shear and intrinsic ellipticity (GI correlations). This measurement uses $36\,278$ Luminous Red 
Galaxies (LRGs) from the Sloan Digital Sky Survey (SDSS) spectroscopic sample with $0.15<z<0.35$, split by redshift and luminosity; $7\,758$ LRGs from 
the 2dF-SDSS LRG and QSO (2SLAQ) Survey at $0.4<z<0.8$; and a variety of other SDSS samples from previous, related work.  We find $>3\sigma$ detections 
of the effect on large scales (up to $60$ \hmpc) for all galaxy subsamples within the SDSS LRG sample; for the 2SLAQ sample, we find a $2\sigma$ 
detection for a bright subsample, and no detection for a fainter subsample.  Fitting formulae are provided for the scaling of the GI correlations with 
luminosity, transverse separation, and redshift (for which the 2SLAQ sample, while small, provides crucial constraints due to its longer baseline in 
redshift).  We estimate contamination in the measurement of $\sigma_8$ for future cosmic shear surveys on the basis of 
the fitted dependence of GI correlations on galaxy properties.  We find contamination to the power spectrum ranging from $-1.5$ (optimistic) to 
$-33$ per cent (pessimistic) for a toy cosmic shear survey using all galaxies to a depth of $R=24$ using scales $l\approx 500$, though the central 
value of predicted contamination is $-6.5$ per cent.  This corresponds to a bias in $\sigma_8$ of $\Delta\sigma_8=-0.004$ (optimistic), $-0.02$ 
(central), or $-0.10$ (pessimistic). We provide a prescription for inclusion of this error in cosmological parameter estimation codes.
The principal uncertainty is in the treatment of the $L\le L_\star$ blue galaxies, for which we have no 
detection of the GI signal, but which could dominate the GI contamination if their GI amplitude is near our upper limits.  Characterization of the 
tidal alignments of these galaxies, especially at redshifts relevant for cosmic shear, should be a high priority for the cosmic shear community.
\end{abstract}


\section{Introduction}

Weak gravitational lensing has emerged in recent years as a powerful tool 
for probing the distribution of matter in the universe (see e.g. the 
review by \citealt{2003ARA&A..41..645R}).  Lensing directly traces the 
matter distribution, and thus is less sensitive to modeling of baryonic 
physics than other cosmological probes.  Because of the relative cleanness 
of the theory, several groups are engaged in a programme to measure 
cosmological parameters using the two-point function of weak lensing 
shear.  The field has grown rapidly, with the first detections 
\citep{2000MNRAS.318..625B, 2000astro.ph..3338K, 2000A&A...358...30V, 
2000Natur.405..143W} giving way to more robust results with smaller statistical 
errors and better control of observational systematics 
\citep{2003MNRAS.341..100B, 2003AJ....125.1014J, 
2005MNRAS.361..160H, 2005MNRAS.359.1277M}.  Several of the more recent
results provide competitive cosmological constraints \citep{2005astro.ph.11089H,
2006ApJ...644...71J, 2006A&A...452...51S}.  

In principle, because the cosmic shear signal has the power to
constrain the matter power spectrum amplitude at a particular
redshift, it may be a powerful tool to constrain models for dark
energy.   Splitting source samples into redshift slices, and auto- and
cross-correlating each slice against itself and against the other
slices can constrain the growth of perturbations as a function of
time.  Indeed, lensing tomography was shown to put significant 
constraints on the dark energy parameters \citep{2002PhRvD..65b3003H, 
2002PhRvD..65f3001H, 2005astro.ph..1594I, 2005PhRvD..71b4026S} and to 
have the potential to test gravitational physics on very large 
scales \citep{2005astro.ph..7184I, 2006PhRvD..74b3512K}.
Therefore, many future surveys are being planned to have sufficiently 
high statistical power to measure dark energy using cosmic shear.  

The cosmological weak lensing signal is small, with typical shear of order 
$10^{-2}$ or less, and some of the more ambitious projects under 
consideration will require measurement of the signal to very high 
fractional accuracy ($<1$ per cent).  This means that even very small 
systematic errors can have a significant influence on cosmic shear 
analyses.  One of these systematics is the problem of measuring shape in 
the presence of an instrumental point spread function (PSF), which can 
easily introduce spurious ellipticity correlations between different 
galaxies and dilute existing correlations by making the galaxies appear 
rounder.  There is now a substantial literature on the subject of 
PSF-induced systematics \citep{2000ApJ...537..555K, 2001A&A...366..717E, 
2002AJ....123..583B, 2003MNRAS.343..459H,  
2004PhRvD..69h3514I, 2004ApJ...613L...1V, 2005A&A...429...75V}, including a detailed comparison 
of different measurement algorithms \citep{2006MNRAS.368.1323H,2006astro.ph..8643M}.  A second 
problem is the determination of the redshift distribution of source galaxies, 
$\rmd N/\rmd z$, which enters into the equation for the cosmic shear power 
spectrum.  There are several ways in which this can be addressed, such as 
direct spectroscopic measurement of some fields 
\citep{2004ApJ...600...17B, 2005PhRvD..71b3002I, 2005MNRAS.361.1287M}, 
self-calibration \citep{2006MNRAS.366..101H}, and possibly in the future 
by large-scale surveys in the H{\sc ~i} 21 cm line 
\citep{2004NewAR..48.1013R}.  Yet a third potential difficulty is the 
uncertainty in the theory: the cosmic shear programme places very 
demanding requirements on the accuracy of $N$-body simulations, and the 
effects of baryons will be important on small scales 
\citep{2004APh....22..211W, 2004ApJ...616L..75Z, 2006ApJ...640L.119J}.

Most of the above-described systematics are technical in nature, and in 
principle can be overcome or at least suppressed with improved modeling 
and data reduction techniques.  Unfortunately, there is another possible 
systematic in weak lensing, namely intrinsic alignments of the source 
galaxy population.  It has been recognized for some time that if the 
intrinsic ellipticities of different galaxies are correlated, then this 
can result in a spurious contribution to the ``shear'' power spectrum. 
Several theoretical models were constructed \citep{2000ApJ...545..561C, 
2000MNRAS.319..649H, 2001MNRAS.320L...7C, 2001ApJ...559..552C, 
2002MNRAS.335L..89J} and the intrinsic alignment correlation function was 
measured by several authors using low-redshift surveys 
\citep{2002MNRAS.333..501B, 2004MNRAS.347..895H}.  
The severity of this problem is lessened by the fact that such
intrinsic ellipticity-intrinsic ellipticity (II)  
correlations can only exist if the two source galaxies are near each other 
in physical space, whereas the shape correlations induced by cosmic shear 
exist between any two galaxies that are near each other in angular 
coordinates, even if they have a large line-of-sight separation.  Thus 
several authors proposed that galaxy pairs at neighbouring redshifts be 
down-weighted or removed in cosmic shear analyses 
\citep{2002A&A...396..411K, 2003A&A...398...23K, 2003MNRAS.339..711H, 
2004ApJ...601L...1T}.  However, it was later found that cosmic shear is 
subject to another type of intrinsic alignment contamination, in which the 
intrinsic ellipticity of a nearby galaxy correlates with the quadrupole of 
the density field surrounding it.  Since this density field produces a 
lensing shear on more distant galaxies, it is possible to have a 
gravitational lensing-intrinsic ellipticity (GI) correlation.  The GI 
correlation can in principle exist between galaxies at very different 
redshifts, and therefore it cannot be eliminated by considering only shear 
cross-correlations between different redshift slices 
\citep{2004PhRvD..70f3526H}.

The cosmic shear community has proposed several methods for addressing the 
GI problem, such as marginalization over parameterized models 
\citep{2005A&A...441...47K, 2007arXiv0705.0166B} and removal based on the redshift dependence 
of the signal \citep{2004PhRvD..70f3526H}.  At the same time, it would be 
useful to know how large a GI signal to expect, including dependence on 
the redshift and on the source galaxy luminosity, colour, environment, 
etc.  Models of GI correlation, either empirical or theoretical, would 
provide an indication of how much of the contaminant needs to be removed 
in order to realize the full potential of cosmic shear, and would help 
guide removal strategies (e.g. rejecting certain types of galaxies with 
the worst GI contamination).  At present, there are several analytical 
\citep{2002astro.ph..5512H, 2004PhRvD..70f3526H} and simulation-based 
\citep{2006MNRAS.371..750H} models, while observationally the GI 
correlation function relevant for lensing has been measured only by 
\citet{2006MNRAS.367..611M}.  There is a much larger literature on galaxy 
alignments that uses statistics other than the GI correlation function, or 
measures correlations at very small scales \citep[for example,
][]{2000ApJ...543L.107P,2001ApJ...555..106L,2002AJ....124..733B,2002ApJ...567L.111L,2004MNRAS.353..529H,2004ApJ...613L..41N,2006ApJ...644L..25A,2006astro.ph..7139A,2006MNRAS.369..479D,2006MNRAS.369.1293Y}. 
These studies  are useful for constraining theoretical models, but
cannot readily be turned into quantitative empirical estimates of
contamination to cosmic shear on large scales.

In a previous paper \citep{2006MNRAS.367..611M} we obtained a detection of 
the large-scale density-ellipticity correlation and presented a 
preliminary analysis of the expected contamination to cosmic shear 
results.  This paper extends that original work by providing a 
phenomenological characterization of the intrinsic alignment 2-point 
correlation functions that is as comprehensive as can be obtained with 
existing data. 
We use a variety of galaxy samples from the SDSS Main spectroscopic sample,
the SDSS spectroscopic Luminous Red Galaxy (LRG) sample, and the 2dF-SDSS
LRG and QSO (2SLAQ) Survey.
In particular, we explore the luminosity, colour, 
environment\footnote{Technically the ``environment dependence of the 
2-point function'' is some combination of higher-order statistics, not a 2-point 
function.  However it can be treated by the same methods used to 
understand colour or luminosity dependence, so we investigate it here.}, 
and redshift dependence of the signal.  This is an important step in 
constructing and validating models (either empirical or simulation-based) 
of intrinsic alignments.  In this paper we will obtain scaling relations 
and construct an empirical model for the intrinsic alignment 2-point 
functions for red galaxies; comparison to simulations and consideration of 
3-point functions will be addressed in a future work.

We begin in Section~\ref{S:formalism} by reviewing briefly the formalism for
describing the cosmic shear and intrinsic alignment correlation
functions.  The sources of data used for this work are described in
Section~\ref{S:data}.  The methodology used for the measurement is
detailed in Section~\ref{S:methodology}. The results obtained and
systematics tests are discussed in Section~\ref{S:results}.
The bias of the LRG samples, needed to estimate the density, is estimated in Section~\ref{S:bias}.
We present power-law fits for the GI correlations
in Section~\ref{S:fits} and compare against theoretical predictions in
Section~\ref{S:theory}. Using these fitting
formulae, we estimate projected contamination for measurements of
$\sigma_8$ with current and future cosmic shear surveys in
Section~\ref{S:contam}; we discuss how cosmic shear analyses can marginalize over our estimates of GI contamination.
We conclude in Section~\ref{S:conclusions}.  Two appendices are included:
Appendix~\ref{app:data} provides our correlation function data, and
Appendix~\ref{app:heymans} provides the calculations for converting the 
\citet{2006MNRAS.371..750H} simulation results to the variables used in this paper.

\section{Formalism}\label{S:formalism}

The formalism for the analysis of the lensing shear two-point function 
\citep{1991ApJ...380....1M} and the intrinsic alignment contamination is 
well developed.  We will briefly summarize the important equations here, 
and then define some new variables that relate to observables in galaxy 
surveys.  Our notation is consistent with that of 
\citet{2004PhRvD..70f3526H} and \citet{2006MNRAS.367..611M}.

\subsection{Correlation functions}

The observed shear $\bgamma$ of a galaxy is a sum of two components: the 
gravitational lensing-induced shear $\bgamma^G$, and the ``intrinsic 
shear'' $\bgamma^I$, which includes any non-lensing shear, typically
due to local tidal fields.  Therefore the $E$-mode shear power
spectrum between  
any two redshift bins $\alpha$ and $\beta$ is the sum of the gravitational 
lensing power spectrum (GG), the intrinsic-intrinsic, and the 
gravitational-intrinsic terms,
\beqa
C_l^{EE}(\alpha\beta) &=&
C_l^{EE,GG}(\alpha\beta)+
C_l^{EE,II}(\alpha\beta)
\nonumber \\ &&
+C_l^{EE,GI}(\alpha\beta).
\eeqa
In \citet{2006MNRAS.367..611M} we presented the Limber integrals that allow
us to determine each of these quantities in terms of the matter
power spectrum and intrinsic alignments power spectrum.  Of greatest interest to us is the GI contamination term,
\beqa
C_l^{EE,GI}(\alpha\beta) \!\!&=&\!\! \int_0^{r_{\rm H}} \frac{\rmd r}{r^2} f_\alpha(r)W_\beta(r)P_{\delta,\tilde\bgamma^I}\left(k=\frac{l+1/2}r\right)
\nonumber \\ &&\!\!
  + (\alpha\leftrightarrow\beta),
\eeqa
where $r_{\rm H}$ is the comoving distance to the horizon, $f_\alpha(r)$ is the comoving distance distribution of the galaxies in sample $\alpha$, and
\beq
W_\alpha(r) = \frac32\Omega_mH_0^2(1+z)\int_r^{r_{\rm H}} \frac{r(r'-r)}{r'} f_\alpha(r') \rmd r'.
\eeq
(The generalization of these equations to curved universes can be found in \citealt{2006MNRAS.367..611M}.)
The power spectrum $P_{\delta,\tilde\bgamma^I}(k)$ is defined as follows.
If one chooses any two points in the SDSS survey, their separation in 
redshift space can then be identified by the transverse separation 
$r_p$ and the radial redshift space separation $\rpi$.\footnote{The 
redshift space separation is frequently denoted $\pi$; we use $\rpi$ to 
avoid confusion since the number $\pi=3.14...$ appears frequently in this 
paper.}  The $+$ and $\times$ components of the shear are measured with 
respect to the axis connecting the two galaxies (i.e. positive $+$ shear 
is radial, whereas negative $+$ shear is tangential).  Then one can write
the density-intrinsic shear correlation in Fourier space as
\beq
P_{\delta,\tilde\bgamma^I}(k) = -2\pi \int \xi_{\delta +}(r_p,\rpi)
J_2(kr_p)
\,r_p\,\rmd r_p\,\rmd\rpi,
\label{eq:j2}
\eeq
where $\xi_{\delta +}(r_p,\rpi)$ is the correlation function between 
$\delta$ and $\tilde\gamma^I_+$.  Here $\tilde\bgamma^I$ is the intrinsic shear weighted by the density of galaxies, i.e. 
$\tilde\bgamma^I=(1+\delta_g)\bgamma^I$, and $\delta$ and $\delta_g$ refer to the matter and galaxy overdensities, respectively.
It is often convenient to do the projection through the radial direction,
\beq
w_{\delta +}(r_p) = \int  \xi_{\delta +}(r_p,\rpi)\rmd\rpi.
\eeq
A similar set of equations can be written for the intrinsic-intrinsic terms; see e.g. Eqs. (5), (7), (9), and (10) of \citet{2006MNRAS.367..611M}.

\subsection{Cosmology dependence}

Here we note the cosmological model and units adopted for this work and the effect of changing it.
Pair separations are measured in comoving $h^{-1}\,$Mpc (where $H_0=100h\,$km$\,$s$^{-1}\,$Mpc$^{-1}$), with the
angular diameter distance computed in a spatially flat $\Lambda$CDM cosmology with $\Omega_m=0.3$.
The correlation function measurement depends on $\Omega_m$ via the determination of both $r_p$ and $\rpi$.
If $\Omega_m\neq 0.3$, then our calculated $r_p$ and $\rpi$ are in error via the equations
\beq
\frac{r_p({\rm calc})}{r_p({\rm true})} = \frac{D_A(z;\Omega_m)}{D_A(z;0.3)}
{\rm ~~and~~}
\frac{\rpi({\rm calc})}{\rpi({\rm true})} = \frac{H(z;0.3)}{H(z;\Omega_m)}.
\eeq
Note that since angular diameter distance $D_A$ is measured in $h^{-1}\,$Mpc and Hubble rate $H(z)$ is measured in 
km$\,$s$^{-1}\,(h^{-1}\,$Mpc$)^{-1}$, there is no dependence of these equations on $h$.  The errors on $r_p$ and $\rpi$ are simple proportionalities 
and hence when we do power-law fits to the projected correlation function, $w_{g+}(r_p)=Ar_p^\alpha$, there is no effect on the 
power-law index $\alpha$.  The amplitude will however be in error by
\beq
\frac{A({\rm calc})}{A({\rm true})} = \frac{H(z;0.3)}{H(z;\Omega_m)}\left[\frac{D_A(z;\Omega_m)}{D_A(z;0.3)}\right]^\alpha.
\label{eq:rescale}
\eeq
This equation should be used when interpreting our results in other cosmologies.  It should be noted however that the effect is rather small:
for $\alpha=-0.8$ (roughly what we observe) and $z=0.6$ (the cosmology matters most at high redshift), the ratio $A({\rm calc})/A({\rm true})=1.05$ for 
$\Omega_m=0.2$ and $0.96$ for $\Omega_m=0.4$.  Since this encompasses the range of $\Omega_m$ values from recent determinations (e.g. 
\citealt{2006astro.ph..3449S}),
the error in our results due to uncertainty in $\Omega_m$ is thus at the $\sim 5$ per cent level, 
which is negligible compared to our final errors.  Therefore we have not included it explicitly in our error analysis, and have not attempted to 
re-measure the correlation function for different cosmologies.

\section{Data}\label{S:data}

Much of the data used here are obtained from the SDSS.  The SDSS 
\citep{2000AJ....120.1579Y} is an ongoing survey to image
$\sim\pi$ steradians of the sky, and follow up $\sim 10^6$ of
the detected objects spectroscopically
\citep{2001AJ....122.2267E,2002AJ....124.1810S,  
2002AJ....123.2945R}. The imaging is carried out by drift-scanning the sky 
in photometric conditions \citep{2001AJ....122.2129H, 
2004AN....325..583I}, in five bands ($ugriz$) \citep{1996AJ....111.1748F, 
2002AJ....123.2121S} using a specially designed wide-field camera 
\citep{1998AJ....116.3040G}. These imaging data are the source of the LSS 
sample that we use in this paper. In addition, objects are targeted for 
spectroscopy using these data \citep{2003AJ....125.2276B} and are observed 
with a 640-fibre spectrograph on the same telescope
\citep{2006AJ....131.2332G}. All of these data are
processed by completely automated pipelines that detect and measure 
photometric properties of objects, and astrometrically calibrate the data 
\citep{2001adass..10..269L, 2003AJ....125.1559P,2006AN....327..821T}. The SDSS
has had six major data releases \citep{2002AJ....123..485S, 
2003AJ....126.2081A, 2004AJ....128..502A, 2005AJ....129.1755A, 
2004AJ....128.2577F, 2006ApJS..162...38A, dr5paper}.  This analysis uses the 
spectroscopically observed galaxies in the Value-Added Galaxy Catalog, 
LSS sample 14 (VAGC; \citealt{2005AJ....129.2562B}), comprising 3423 square 
degrees with SDSS spectroscopic coverage, as well as photometric
catalogs to be described below covering a larger area.

To determine density-shear correlation functions, one
needs a sample of galaxies that traces the intrinsic shear, and
another that traces the density field.  Unlike in
\citet{2006MNRAS.367..611M}, for this paper we use a variety 
of galaxy samples spanning redshifts $z=0.05-0.8$.  The sources of
these samples are the SDSS Main spectroscopic sample ($z=0.05-0.2$),
the SDSS spectroscopic LRG sample
($z=0.16-0.36$), and the 2SLAQ spectroscopic LRG sample ($z=0.4-0.8$).
In the following
subsections we describe each of these samples, and
Table~\ref{T:samples} summarizes their characteristics.  In each of these three redshift regions, the full sample is used to trace the density field, 
but we break the galaxies down into subsamples based on luminosity and (for the Main sample) colour to trace the intrinsic shear field.  This enables 
us to explore the possibility that different types of galaxies show different intrinsic alignment signals.

\begin{table*}
\caption{\label{T:samples}Description of different galaxy samples
and their subsamples used for the intrinsic alignment measurements.
Included is a sample code and a brief description of the sample, the
redshift range spanned, the absolute magnitude and colour cuts (including the redshift to which the absolute magnitude was $k+e$-corrected),
the magnitude type (P for Petrosian, M for model), and the number of galaxies.}
\begin{tabular}{crccrr}
\hline\hline
Sample & Sample description & $z$ range & $M_r$ cuts & $m_r$ type & $N_{gal}$ \\
\hline
L3 & SDSS Main L3 (Section~\ref{SSS:main}) & $0.02<z<0.12$ & $-20
\le M_r^{0.1} \le -19$ & P & $66\,312$ \\
L4 & SDSS Main L4 (Section~\ref{SSS:main}) & $0.02<z<0.19$ & $-21
\le M_r^{0.1} \le -20$ & P & $118\,618$ \\
L5 & SDSS Main L5 (Section~\ref{SSS:main}) & $0.02<z<0.29$ & $-22 \le M_r^{0.1}
\le -21$ & P & $73\,041$ \\
L6 & SDSS Main L6 (Section~\ref{SSS:main}) & $0.03<z<0.35$ & $-23 \le M_r^{0.1}
\le -22$ & P & $7\,937$ \\
\hline
LRG & SDSS LRG (Section~\ref{SSS:specLRG})$\qquad\qquad$ &
$0.16<z<0.35$ &  & M & $36\,278$ \\
LRG1 & cut by magnitude & & $M_r^{0.0}\ge -22.3$ & & $16\,068$ \\
LRG2 & cut by magnitude & & $-22.6\le M_r^{0.0}<-22.3$ & & $13\,019$ \\
LRG3 & cut by magnitude & & $M_r^{0.0}<-22.6$ & & $7\,191$ \\
LRG.\BCG & cut by environment & & \BCG\ only & & $30\,849$ \\
LRG.\nonBCG & cut by environment & & \nonBCG\ only & & $5\,429$ \\
\hline
2SLAQ & Full 2SLAQ LRG (Section~\ref{SSS:2SLAQ}) & $0.4<z<0.8$ & & M & $7\,758$ \\
2SLAQf & cut by magnitude & & $M_r^{0.0} > -22$ & M & $3\,768$ \\
2SLAQb & cut by magnitude & & $M_r^{0.0} < -22$ & M & $3\,490$ \\
\hline\hline
\end{tabular}
\end{table*}

\subsection{SDSS Main subsamples}\label{SSS:main}

The first samples of galaxies we discuss are those used in
\citet{2006MNRAS.367..611M}, subsamples of the SDSS Main spectroscopic
sample divided by luminosity and other properties.  The GI correlation models in this paper are split into early and late types, so we have presented 
here the GI signals for the Main galaxies split both by luminosity, and by colour as ``blue'' or ``red.''  (The
differences between the Main sample analysis of this section and that of \citealt{2006MNRAS.367..611M} are the inclusion of the colour separator, and 
the use of 
the full Main sample rather than the luminosity subsamples to trace the density field.)
For this work, we
use the galaxies in L3 through L6, one magnitude fainter than $L_*$ (L3)
through two magnitudes brighter (L6).  The sample properties were described in full in that
paper; for this work, we mention only that the luminosities described
are Petrosian magnitudes, extinction corrected using reddening maps
from \citet{1998ApJ...500..525S} with the extinction-to-reddening
ratios given in \citet{2002AJ....123..485S}, and $k$-corrected to $z=0.1$ using
{\sc kcorrect v3\_2} software as described by \citet{2003AJ....125.2348B}.

We use an empirically-determined
redshift-dependent colour separator of $u-r = 2.1 + 4.2z$, where we use
observer frame rather than rest-frame colours; within these luminosity
bins, the fraction of red galaxies is 0.40 (L3), 0.52 (L4), 0.64 (L5),
and 0.80 (L6). 

\subsection{SDSS spectroscopic LRGs}\label{SSS:specLRG}

Another sample we use is the DR4 spectroscopic LRG sample
\citep{2001AJ....122.2267E}.  This sample has a fainter flux limit
($r<19$) than the Main sample, 
and colour cuts to isolate LRGs.  We include these galaxies in the
redshift range 
$0.16<z<0.35$, for which the sample is approximately volume-limited
and includes 36~278 galaxies total.

In order to study variation within this sample, we use cuts on several
parameters.  First, we construct luminosities using the $r$-band
model magnitudes, and define three luminosity subsamples as shown in
Table~\ref{T:samples}.  The absolute magnitude cuts 
are defined in terms of $h=H_0/(100\,$km$\,$s$^{-1}\,$Mpc$^{-1})$ such 
that one can implement the cuts without specifying the value of $H_0$.  
The magnitudes have been corrected for extinction, and are
$k+e$-corrected to $z=0$ using the same templates as in \cite{2006MNRAS.372..537W}.
We exclude galaxies lying inside the bright   
star mask.  Random catalogs were generated taking into account the 
variation of spectroscopic completeness with position. The random points 
were assigned absolute magnitudes and redshifts drawn from a
distribution consistent  
with the real sample. Shape measurements were obtained via  
re-Gaussianization for 96 per cent of this sample (see
\S\ref{SS:ellipdata}); the  
remainder failed due to various problems, such as interpolated or saturated 
centers.

Besides luminosities, we also use measures of local environment to understand variation of intrinsic alignments within the sample.  In particular, one 
question we may ask is whether intrinsic alignments are properties only of the central galaxy in a halo, or if satellite galaxies have intrinsic 
alignments as well.  Unfortunately it is very difficult to actually measure which galaxies are ``central.'' A related quantity that we can measure is 
whether the galaxy is a brightest group/cluster galaxy (\BCG) or \nonBCG.  To isolate \BCG s, we require that a given LRG be the brightest 
spectroscopic LRG within 2 \hmpc\ projected separation and $\pm 1200$ km~s$^{-1}$.  This cut designates 83 per cent of the sample as \BCG s and 17 per 
cent as \nonBCG s; the ``\BCG s'' are either in the field or host groups and clusters.  One limitation to this cut is fiber collisions, for which the 
relevant radius is $55''$ (corresponding to $\sim 200$ \hkpc\ at the range of redshifts of the sample), which may cause us to mistakenly identify 
\nonBCG s as \BCG s.  This is, however, only relevant on small scales.  We emphasize that this distinction between \BCG s and \nonBCG s does not place 
any significant constraint on the ``\nonBCG'' galaxy luminosity, which may be quite close to the \BCG\ luminosity; the median luminosity gap is 0.3 
magnitudes.  Unfortunately, the sample of \nonBCG s is too small to obtain meaningful results when splitting by luminosity gap.  It is encouraging that 
our ``\nonBCG'' fraction of 0.17 agrees with the satellite fraction estimated from halo modeling of the LRG three-point function \citep{kulkarni06}.  
However, in interpreting the results, it must still be remembered that there could be cases where the \BCG\ is not actually the central galaxy (in some 
cases, such as merging clusters, there may not even be such a thing as a ``central'' galaxy), and that our selection is restricted to LRGs.  In 
principle one could have increased the signal-to-noise ratio for the \nonBCG s by increasing the transverse separation cut to $>2$\hmpc\ so as to 
obtain equal numbers in the \BCG\ and \nonBCG\ samples, but this would have had the undesired effect of cutting on large-scale environment (i.e. far 
beyond the virial radius), thereby including many central galaxies in the ``\nonBCG'' sample.

Finally, in order to constrain redshift evolution more tightly, we have
split each of the three LRG luminosity samples at $z=0.27$ to create a
total of six spectroscopic LRG samples.

\subsection{2SLAQ LRGs}\label{SSS:2SLAQ}

The 2dF-SDSS LRG and QSO (2SLAQ) Survey has produced samples of
roughly 10~000 LRGs with $z_{med} \sim 0.55$ with spectroscopic
redshifts, and 10~000 faint quasars.  For this paper, we use the 2SLAQ
LRGs, which have similar cuts to the SDSS spectroscopic LRG sample,
with a fainter apparent magnitude cut of $m_i(\mbox{model}) <
19.8$ and with a slightly bluer rest-frame colour cut $c_{\perp}$ than
in the SDSS LRG selection.

\cite{2006MNRAS.372..425C} includes further details of the 2SLAQ
selection and observation; \cite{2006MNRAS.372..537W} shows a first
analysis of the LRG luminosity function evolution using this sample
in comparison with the SDSS spectroscopic LRG sample.  While roughly
10~000 LRG redshifts have been measured, several of our criteria
reduce the sample size: we require that they pass the cuts to be in
the LRG primary sample (\citealt{2006MNRAS.372..537W}, sample 8) rather than secondary sample (sample 9, which
has far lower completeness); we eliminate repeat observations; we
reject those in fields with significant incompleteness; we reject
those with poor redshift quality; we require that those with
redshifts lie in the range $0.4<z<0.8$; we require that they have
shape measurements with sufficiently high resolution factor
(\S\ref{SS:ellipdata}).  The last of these cuts reduces the final
sample from the canonical 8~656 to 7~758 galaxies.
Random catalogs with the same completeness as a function of angular
and radial position as the real sample were used for the random
points.

Due to the faintness of this sample, we use model magnitudes.  These
are $k+e$-corrected to $z=0$ using predictions derived from
\citet{2003MNRAS.344.1000B}; the limitations of these models in
describing this sample are discussed more fully in \cite{2006MNRAS.372..537W}.

\subsection{Ellipticity data}\label{SS:ellipdata}

In addition to a sample of galaxies, we also need their ellipticities.  
For this purpose, we use the measurements by \citet{2005MNRAS.361.1287M}, who 
obtained shapes for more than 30 million galaxies in the SDSS imaging data 
down to extinction-corrected magnitude $r=21.8$, (i.e. far fainter than the spectroscopic limit 
of the SDSS).  A minor modification to the {\sc Reglens} pipeline as described in
\citet{2006MNRAS.367..611M} was also used.

The {\sc Reglens} pipeline obtains galaxy images in the $r$ 
and $i$ filters from the SDSS ``atlas images'' 
\citep{2002AJ....123..485S}.  The basic principle of shear measurement 
using these images is to fit a Gaussian profile with elliptical isophotes 
to the image, and define the components of the ellipticity
\beq
(e_+,e_\times) = \frac{1-(b/a)^2}{1+(b/a)^2}(\cos 2\phi, \sin 2\phi),
\label{eq:e}
\eeq
where $b/a$ is the axis ratio and $\phi$ is the position angle of the 
major axis.  The ellipticity is then an estimator for the shear,
\beq
(\gamma_+,\gamma_\times) = \frac{1}{2\cal R}
\langle(e_+,e_\times)\rangle,
\eeq
where ${\cal R}\approx 0.87$ is called the ``shear responsivity'' and 
represents the response of the ellipticity (Eq.~\ref{eq:e}) to a small 
shear \citep{1995ApJ...449..460K, 2002AJ....123..583B}.  In practice, a 
number of corrections need to be applied to obtain the ellipticity.  The 
most important of these is the correction for the smearing and 
circularization of the galactic images by the PSF; 
\citet{2005MNRAS.361.1287M} uses the PSF maps obtained from stellar images 
by the {\sc psp} pipeline \citep{2001adass..10..269L}, and corrects
for these 
using the re-Gaussianization technique of \citet{2003MNRAS.343..459H},
which includes corrections for non-Gaussianity of both the galaxy
profile and the PSF.  A 
smaller correction is for the optical distortions in the telescope: 
ideally the mapping from the sky to the CCD is shape-preserving 
(conformal), but in reality this is not the case, resulting in a nonzero 
``camera shear.'' In the SDSS, this is a small effect (of order 0.1
per cent) which can be identified and removed using the astrometric solution 
\citep{2003AJ....125.1559P}.  Finally, a variety of systematics tests are 
necessary to determine that the shear responsivity ${\cal R}$ has in fact 
been determined correctly.  We refer the interested reader to 
\citet{2005MNRAS.361.1287M} for the details of these corrections and 
tests.

\section{Methodology}\label{S:methodology}

The basic methodology in this paper is very similar to that in 
\cite{2006MNRAS.367..611M}: we
correlate a sample of galaxies against some subset of itself, using the subset to trace
intrinsic alignments and the full sample to trace the density field.  
We use the same set
of estimators as in \cite{2006MNRAS.367..611M}, generalizing the LS
\citep{1993ApJ...412...64L} estimator commonly used for galaxy-galaxy
autocorrelations.

The generalization of this estimator to the galaxy-intrinsic
correlation is 
\beq
\hat\xi_{g+}(r_p,\rpi) = \frac{S_+(D-R)}{RR} = \frac{S_+D-S_+R}{RR},
\label{eq:lsxids}
\eeq
where $S_+D$ is the sum over all pairs with separations $r_p$ and $\rpi$ 
of the $+$ component of shear:
\beq\label{E:lsxiss}
S_+D = \sum_{i\neq j| r_p,\rpi} \frac{e_+(j|i)}{2\cal R}, 
\eeq
where $e_+(j|i)$ is the $+$ component of the ellipticity of galaxy $j$ 
measured relative to the direction to galaxy $i$, and ${\cal R}$ is the 
shear responsivity.  $S_+R$ is
defined by a similar equation. We emphasize that positive $\xi_{g+}$ indicates a
tendency to point towards overdensities of galaxies (i.e., radial
alignment, the opposite of the convention in galaxy-galaxy lensing
that positive shear indicates tangential alignment).

We also generalize this large-scale correlation function estimator to
cross-correlations.  In this case, we use one sample to trace the
intrinsic shear, and other to trace the density field.  
Thus in Eq.~(\ref{eq:lsxids}), we find pairs of galaxies such that one is in
the shear sample and the other in the density sample, so that the
$S_+$ is determined from the former, and the $D-R$ from the latter;
the $RR$ in the bottom is determined using one random point
corresponding to the shear sample and the other corresponding to the
density sample. 

This correlation function estimator is then integrated along the line
of sight to form the projected intrinsic shear - density correlation
function $w_{g+}(r_p)$.  We model this function as a power-law,
$w_{g+}=Ar_p^{\alpha}$, where fits for $A$ and $\alpha$ are done using
the full jackknife covariance matrix.  

For this paper, we used the same two software pipelines as in
\citet{2006MNRAS.367..611M} to compute the large-scale density-shear
correlations. We independently generalized both pipelines
to allow different galaxy samples to be used to trace the 
intrinsic shear and the density field. In summary, the pipeline for which we present 
results computes the correlation functions over a $120$ \hmpc\ 
(comoving) range along the line of sight ($-60 < \Pi < +60$ \hmpc) 
divided into 30 bins, then integrates over $\Pi$.  The range of 
transverse separations is from $0.3$ to $60$ \hmpc, in 10 logarithmic 
bins.  Covariance matrices were determined using a jackknife with 50 regions; 
as demonstrated there, results were relatively robust to the number of
bins, with 100 regions giving approximately the same detection
significance as 50 regions. For more details about these pipelines, 
see \citet{2006MNRAS.367..611M}.  

\section{Results}\label{S:results}

In this section we describe the results of the galaxy density-shape correlation functions for the SDSS Main samples, the SDSS LRGs, and the 2SLAQ LRGs.  
The measured correlation functions and their uncertainties and correlation matrices are presented in Appendix~\ref{app:data}.

\subsection{SDSS Main sample}
\label{ss:sdss-main}

\begin{figure}
\includegraphics[width=3.2in]{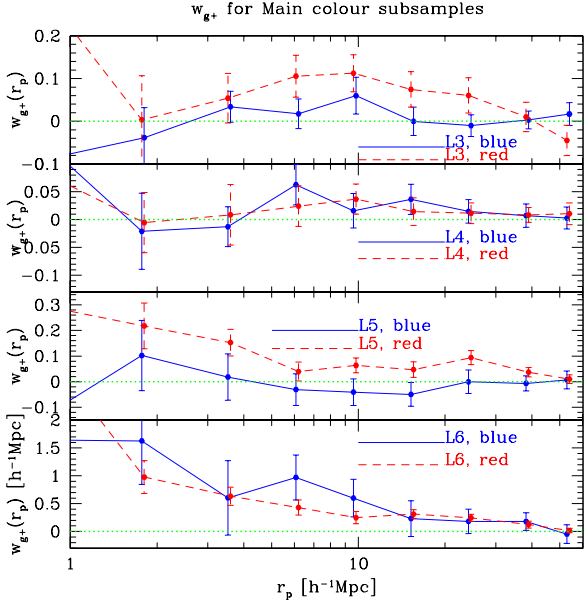}
\caption{\label{fig:main}The GI correlation functions for the SDSS Main subsamples, split into colour and luminosity bins.}
\end{figure}

\begin{figure}
\includegraphics[width=3.2in]{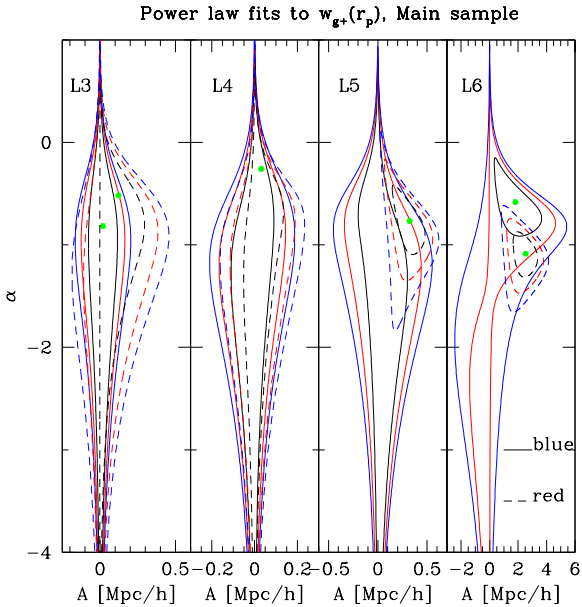}
\caption{\label{F:allcolorconf}Confidence contours for power-law fits to
  $w_{g+}(r_p)$ for Main sample galaxies.  Contours are
  shown for various subsets of 
  data labelled on the plots; in each case, $1\sigma$, $2\sigma$, and
  $3\sigma$ contours are shown.}    
\end{figure}

Here we present results of the measurements of the galaxy
density-shape correlations $w_{g+}(r_p)$ using the Main sample split not only into
luminosity bins, but also into colour samples.  There are 8 subsamples total, since we have 4 luminosity bins and 2 colour bins.
We attempt to address the issue of whether blue galaxies show any density-shape alignment
when correlated against the full L3--L6 (all colours), and to place constraints if there is no
detection.  The GI signal $w_{g+}(r_p)$ for each of the eight subsamples is shown in Figure~\ref{fig:main}.
Figure~\ref{F:allcolorconf} shows the confidence contours for
fits to a power law, $w_{g+}(r_p)=Ar_p^{-\alpha}$ as discussed in \cite{2006MNRAS.367..611M}.

As shown, for both colour subsamples there is hardly any detection in
L3 and L4, consistent with previous works.  There is a hint of a
signal in L3 for the
red sample: when we fit the whole range of scales to an arbitrary
power law there is no detection, however as we will see in
Section~\ref{S:fits} if we  
fit to large scales ($r_p>4.7h^{-1}\,$Mpc) where the bias is expected
to be roughly linear, and restrict to the power law $\alpha=-0.73$
observed for  
the LRGs, there is a marginal ($2.4\sigma$) detection.  For L5, the
detection with the red subsample is robust whereas there is no
detection with the  
blue subsample.  For L6, the constraints are weak with the blue sample
due to its small size, so while the magnitude of the alignments are
consistent  
with the red subsample, they are also marginally consistent with zero.  The
rest-frame colour distribution of the L6 blue sample, and the
distribution of Photo
pipeline output {\tt frac\_deV} (a measure of the degree to which a
galaxy profile is closer to an exponential or de Vaucouleur profile),
suggest that this small L6 blue sample may contain galaxies that are
on the edge of the blue vs. red galaxy distinction, which could 
explain this consistency of results.

\subsection{SDSS LRGs}

\begin{figure}
\includegraphics[width=3.2in]{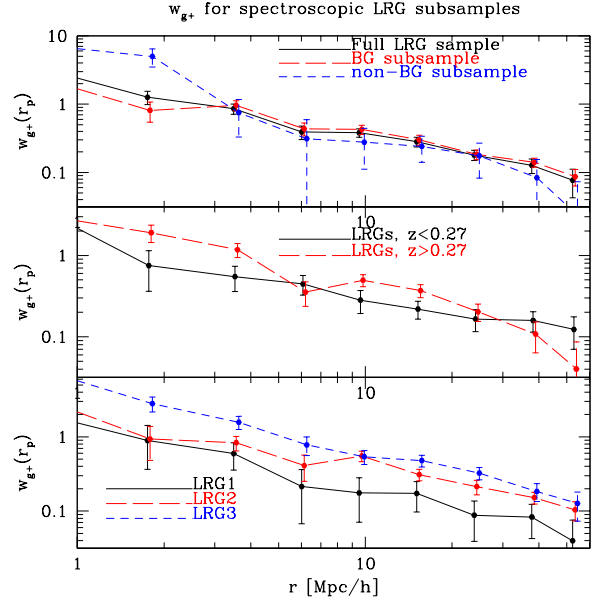}
\caption{\label{F:wgpluslrg}The density-shape correlation function
  $w_{g+}(r_p)$ from $1$--$60$ \hmpc\ with the full spectroscopic LRG
  sample and various subsamples as labeled on the plot.  Errors are
  $1\sigma$ but are somewhat correlated on large scales.}   
\end{figure}

\begin{figure*}
\includegraphics[width=5.5in]{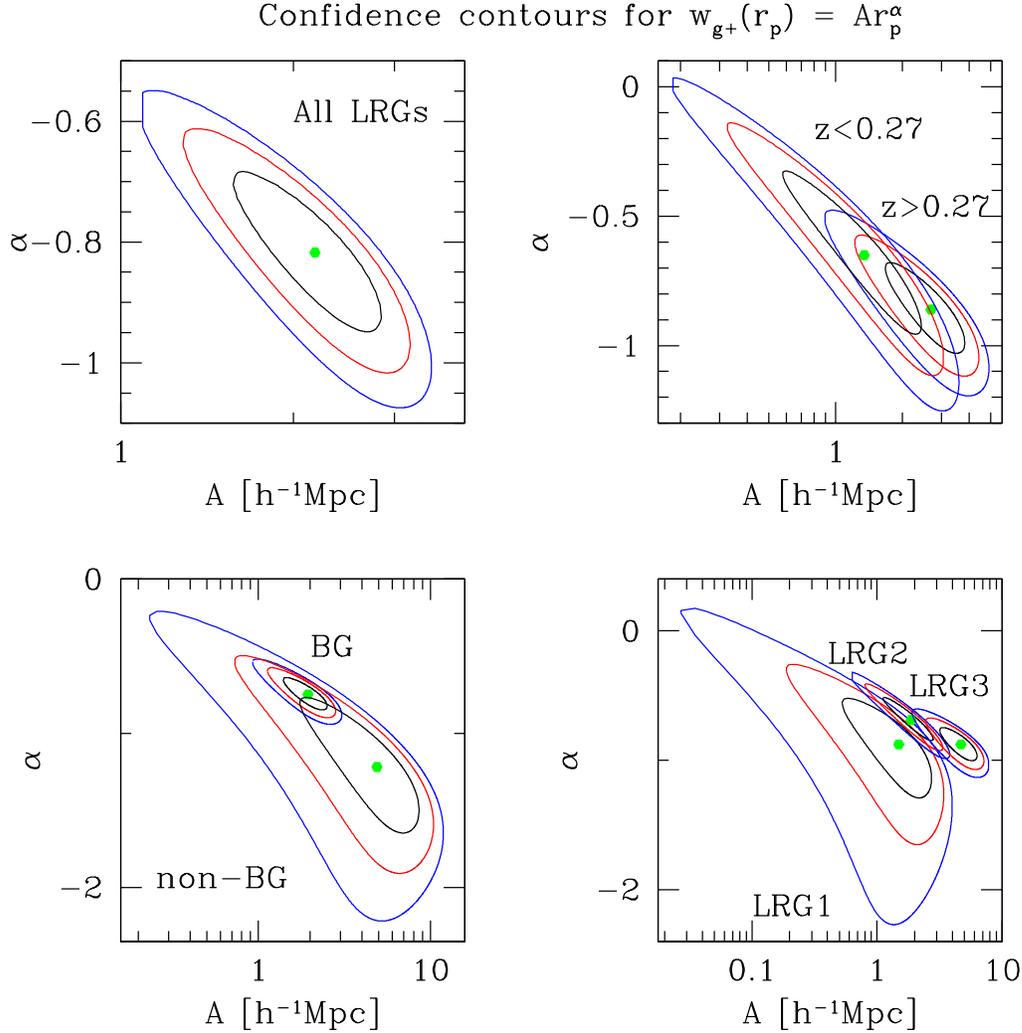}
\caption{\label{F:alllrgconf}Confidence contours for power-law fits to
  $w_{g+}(r_p)$ for SDSS spectroscopic LRGs.  Contours are shown for various subsets of
  data labelled on the plots; in each case, $1\sigma$, $2\sigma$, and
  $3\sigma$ contours are shown.}    
\end{figure*}

Here we present results of the measurements of
density-shape correlations for the SDSS spectroscopic LRGs (the measurement is otherwise similar to that of \citealt{2006MNRAS.367..611M}).  The
plots of $w_{g+}(r_p)$ in Figure~\ref{F:wgpluslrg} are in the same form as
in that paper,  
including $1\sigma$ errors.  Figure~\ref{F:alllrgconf} shows the
confidence contours for fits to a power-law, $w_{g+}=Ar_p^{\alpha}$.

In the top panel of Figure~\ref{F:wgpluslrg}, we show $w_{g+}(r_p)$ for
the full spectroscopic LRG sample, and the \BCG\ and \nonBCG\ 
subsamples.  As shown, the full sample and the \BCG\ subsamples are
robustly detected on all scales, out to $60$ \hmpc.  The \nonBCG\ 
subsample has a significantly lower signal to noise due to its small
size, but the amplitude appears roughly comparable to that of the \BCG\ subsamples.  In Figure~\ref{F:alllrgconf}, the contours for the full LRG
sample are shown in the upper left panel, and for \BCG s and \nonBCG s
separately in the lower left panel.  As shown there, the constraints
on the \nonBCG\ sample are indeed quite weak, but the detection is
still robust at slightly higher than the $3\sigma$ level.  The
amplitude is higher and the power-law steeper than for the \BCG\ 
sample, but is still consistent with it at the $1\sigma$ level.

In the middle panel of Figure~\ref{F:wgpluslrg}, we show $w_{g+}(r_p)$
for the spectroscopic LRG sample split at the median redshift.  The
purpose of this test is that, since this sample is roughly volume
limited, with the same luminosity-distribution in each of the halves,
any evolution in the intrinsic alignment amplitude should be due to
redshift evolution.  While the amplitude appears higher for the higher
redshift sample, the difference is not large; we must see the
confidence contours on the fit parameters (which take into account the
error correlations) before deciding if the difference is
statistically significant, and will also later account for evolution
of bias with redshift in Section~\ref{S:fits}.  These contours appear in the upper right
panel of Figure~\ref{F:alllrgconf}, and indicate that the difference is,
indeed, not statistically significant.

Finally, the bottom panel of Figure~\ref{F:wgpluslrg} shows
$w_{g+}(r_p)$ for the spectroscopic LRG sample split into luminosity
subsamples.  These results appear to have roughly the same power-law
index, with a multiplicative difference in amplitude, and the
confidence contours in the lower right panel of
Figure~\ref{F:alllrgconf} indicate that this is indeed the
case.  

\subsection{2SLAQ LRGs}

In this section, we present results of the measurement of the
density-shape correlation using the 2SLAQ LRG sample.
Unfortunately, we cannot obtain meaningful results by comparing the full 2SLAQ sample 
results against the full spectroscopic LRG sample, because the samples
do not cover the same range of absolute magnitudes.  The 2SLAQ sample is, on average, fainter and 
bluer than the spectroscopic LRG sample.  This difference is
problematic since it will tend to give a lower intrinsic alignment
amplitude for the 2SLAQ sample that we will need to separate from redshift
evolution effects. 

\begin{figure}
\includegraphics[width=3.2in]{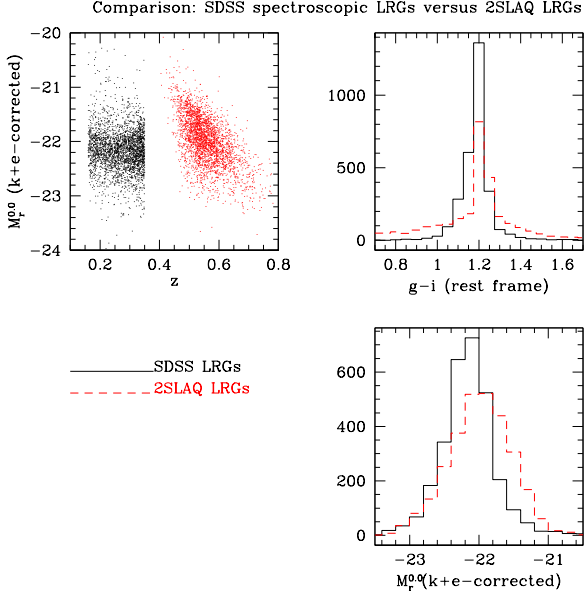}
\caption{\label{F:compabsmag}A comparison of the $r$-band absolute
  magnitudes and colours of the SDSS and 2SLAQ LRG samples.  The upper
  left panel shows the absolute magnitudes as a function of redshift.
  The upper right shows the distribution of $g-i$ rest frame colour,
  and the lower right shows the absolute magnitude histograms over the
  full samples.}   
\end{figure}

 To illustrate this point, Figure~\ref{F:compabsmag}
shows a comparison of the $k+e$-corrected (to $z=0$) model $r$-band absolute
magnitude distributions of the samples.  First, the upper left panel,
which shows scatter plots of magnitude $M_r^{0.0}$ versus redshift
$z$.  As shown, the spectroscopic LRG sample has roughly constant mean
absolute magnitude as a function of redshift for the full redshift
range used here.  The 2SLAQ sample, on the other hand, shows a trend
of being fainter at the low redshift end and brighter at the high
redshift end.  The mean absolute magnitude is roughly $0.2$ magnitudes
fainter for the 2SLAQ sample than for the SDSS spectroscopic LRG
sample (this can also be seen in the bottom right panel, which shows
histograms of the absolute magnitude values).  This difference is a
concern for our work, since in the previous subsection we showed that the intrinsic 
alignment amplitude has a very strong scaling with luminosity.

One might wonder whether this difference that we observe is due to
some systematic, such as uncertainty in the k+e-corrections.
\cite{2006MNRAS.372..537W}  explore the difference in more detail, and it
does not seem possible to design a plausible model with different
formation 
times, levels of star formation, or other differences that would
reduce the difference 
between the samples to zero.  They are inherently different.

\begin{figure}
\includegraphics[width=3.2in]{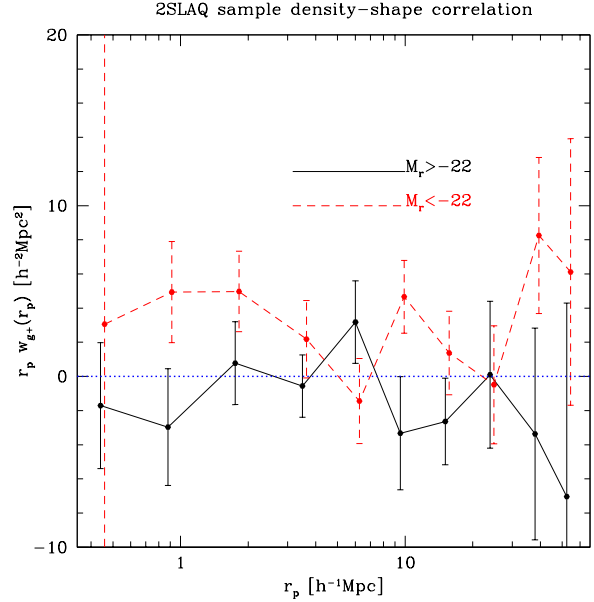}
\caption{\label{F:wgplus2slaq}The galaxy density-shape correlation function
  $r_p w_{g+}(r_p)$ from $1$--$60$ \hmpc\ with the luminosity subsamples used to trace the shapes and the full 2SLAQ
  sample used for the galaxy density $g$.  Errors are
  $1\sigma$ but are slightly correlated on large scales.}   
\end{figure}

\begin{figure}
\includegraphics[width=3.2in]{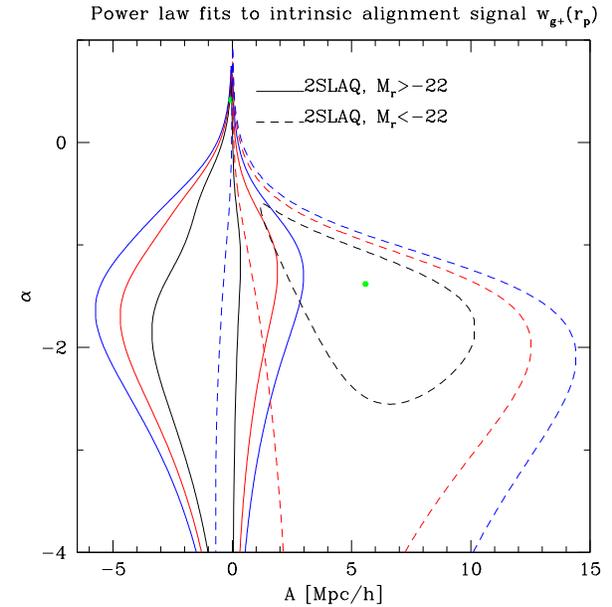}
\caption{\label{F:conf2slaq}Confidence contours for power-law fits to
  $w_{g+}(r_p)$.  $1\sigma$, $2\sigma$, and
  $3\sigma$ contours are shown for both luminosity subsamples.}    
\end{figure}

Because of this difference between the samples, rather than comparing
intrinsic alignment amplitudes directly, we choose instead to fit to a
model of the intrinsic alignment amplitude as a function of luminosity
and redshift separately. This work can be aided by splitting the 2SLAQ
sample into bright and fainter subsets at $M_r=-22$ in addition to
using the six redshift and luminosity samples of SDSS spectroscopic
LRGs.  
Figure~\ref{F:wgplus2slaq} shows $r_p w_{g+}(r_p)$ for the two 2SLAQ
luminosity samples, and Figure~\ref{F:conf2slaq} shows the confidence
contours for a 
fits to a power-law for each subsample.

Results for the luminosity and redshift-dependent fits will be
presented in Section~\ref{S:fits}.

\subsection{Systematics tests}

As in \cite{2006MNRAS.367..611M}, we have done several systematics tests to ensure that the detections have not been contaminated by spurious 
instrumental or other effects.  The first is the standard 45-degree test, whereby we rotate all source ellipticities by 45 degrees before computing the 
correlation functions (to obtain $w_{g\times}$ instead of $w_{g+}$).  This rotated correlation function reverses sign under parity (i.e. the sign of 
the correlation function is flipped depending on whether one rotates clockwise or counterclockwise) and therefore cannot 
appear unless there is a systematic, or galaxy formation violates parity invariance.\footnote{Note that for intrinsic-intrinsic ellpiticity 
correlations the 
rotated correlation function $w_{\times\times}$ is parity-allowed and so this would not be a good systematics test; rather one would use $w_{+\times}$, 
i.e. rotate the ellipticity of only one of the two galaxies in question.}  This test was done for the six subsamples of
the SDSS spectroscopic LRGs (splitting jointly into three luminosity and
two redshift bins), for the ``\BCG'' and ``\nonBCG'' subsamples of
the SDSS LRGs, and for the two luminosity subsamples of the 2SLAQ
LRGs, giving ten $w_{g\times}$ computations total.  For each computed
$w_{g\times}$ we computed the $\chi^2$ for a fit to zero signal, and
found the associated $p$-values (the probability of getting a larger
value of $\chi^2$ by chance).  We note that, as in
\cite{2006MNRAS.367..611M}, the $\chi^2$ values do not follow the
usual $\chi^2$ distribution because of noise in the jackknife
covariance matrices.  The formalism to describe this effect and
account for it with simulations was developed in \cite{2004MNRAS.353..529H}.  
The ten $p(>\chi^2)$-values computed taking into account the modified
distribution of $\chi^2$ ranged from 0.2--0.94; the lack of very low values
indicates that we do not have significant B-mode contamination.

Another test from \cite{2006MNRAS.367..611M} is the integration of the
3-dimensional correlations $\xi_{g+}$ over large line-of-sight
separations, $30<|\Delta\Pi|<90$ \hmpc.  A signal that is of
astrophysical origin should be dominated by the results from smaller
$|\Delta\Pi|$, and hence should be consistent with zero for this
range.  This test will allow us to rule out, e.g., some optical
effect causing an apparent alignment of galaxy images (which would not
depend on the relative line-of-sight separation).  For this test, in
all cases, the signal when integrating over large line-of-sight
separations was consistent with zero, with $p(>\chi^2)$ values ranging from
$0.07$ to $0.94$. 

As in \cite{2006MNRAS.367..611M}, we return to the question of the
size of jackknife subsamples.  This was primarily a concern for the
lower redshift subsamples, for which a jackknife region of a
particular size corresponded to a small comoving transverse
separation.  Nonetheless, despite the higher average redshift of the
samples in this paper, we revisit this issue briefly.  A particular
concern is for the 2SLAQ sample which, while at higher redshift
than any of the other samples, also covers a much smaller area overall,
so independence of the jackknife regions may be a concern.  For the
samples used here, changing the number of jackknife samples by a
factor of two did not change the size of the errors by more than 5-10
per cent, which suggests that we are using the jackknife in a regime
where it has converged.

A final concern we may have in comparing results from the
SDSS spectroscopic LRG at $z\sim 0.3$ and the 2SLAQ LRGs at $z\sim
0.55$ is that the shape measurements, which are averaged over $r$ and
$i$ bands, are for different sets of rest-frame wavelengths.  In
principle, these may then be dominated by different stellar
populations within the same galaxy that have different intrinsic
alignment properties.  To test for this effect, we have calculated
the analogy of the GI correlation function in Eq.~(\ref{eq:lsxids}) for
SDSS LRGs using the difference between $r$- and $i$-band shears rather
than their average as is usually done.  Assuming that intrinsic
alignments do not vary across this range of wavelengths, this quantity
should then be consistent with zero.  
(We remind the reader that
$r$ band for the SDSS LRGs corresponds roughly to $i$ band for the
2SLAQ LRGs when considering the rest-frame wavelengths.)  For the six
SDSS LRG subsamples, these correlation functions were consistent with
zero, with $p(>\chi^2)$ values ranging from 0.29--0.81. 

\section{Bias of density tracers}
\label{S:bias}

To convert the observed $w_{g+}$ to the more relevant (for cosmic shear) $w_{\delta +}$, we need to know
the bias of the sample used as the density tracer.  This needs to be done for the two LRG samples, and for the SDSS Main sample.  Due to the much 
higher signal-to-noise in the LRG samples, we have measured the bias
from the LRG projected autocorrelation function.  The SDSS Main sample has a much lower signal-to-noise detection of GI alignments, and so we have 
constructed a crude bias estimate by combining previously published results.

\subsection{LRGs}

Besides computing $w_{g+}$, our code computes $w_{gg}$, the galaxy-galaxy autocorrelations.  We have compared our results for the full SDSS 
spectroscopic LRG sample with those in \cite{2005ApJ...633..560E} and
\cite{2005ApJ...621...22Z} and found agreement at the $1\sigma$
level.  Likewise, our results for the 2SLAQ sample are consistent with
those in
\cite{rossetal}, though the central value of
derived bias does not precisely agree due to significantly different
analysis methodologies.  We determine consistency by computing the
best-fit powerlaw to $w_{gg}$ and comparing the $\chi^2$ for our
best-fit parameters versus the $\chi^2$ for the best-fit parameters in
\cite{rossetal}, and find $p(>\Delta\chi^2)=0.2$.

The projected autocorrelations are computed for $0.3<r_p<60$ \hmpc{}
using  
\begin{equation}
w_{gg}(r_p) = \int \xi_{gg}(r_p, \Pi) d\Pi,
\end{equation}
where $\xi_{gg}$ has been estimated using the LS estimator, 
\begin{equation}
\hat{\xi}_{gg}(r_p,\Pi) = \frac{(D-R)^2}{RR} = \frac{DD-2DR+RR}{RR}.
\label{eq:lsest}
\end{equation}
Integration along the $\Pi$ direction is carried out for $-60 <
\Pi < +60$ \hmpc.  As for the $w_{g+}$ calculations, covariance
matrices are determined 
using a jackknife with 50 regions.

We did three types of fits to obtain the bias from $w_{gg}$ for each of the four samples (low-$z$ SDSS LRGs, high-$z$ SDSS LRGs, faint 2SLAQ LRGs, and 
bright 2SLAQ LRGs).  The first method is to fit the bias $b$ to the linear correlation function,
\beq
w_{gg}(r_p) = b^2\int
\frac{k\,\rmd k}{2\pi}
P_{\rm lin}(k)W_{r_p}(k)+C.
\label{eq:w-lin}
\eeq
The second method is to use the $Q$-model from \cite{2005MNRAS.362..505C},
\beq
w_{gg}(r_p) = b^2\int
\frac{k\,\rmd k}{2\pi}
P_{\rm lin}(k)
\frac{1+Qk^2}{1+Ak}W_{r_p}(k)+C,
\eeq
where for a real-space power spectrum $A=1.7h^{-1}\,$Mpc \citep{2005MNRAS.362..505C}; the $\chi^2$ is minimized with respect to both $b$ and $Q$.  The 
third method is to use Eq.~(\ref{eq:w-lin}) but with the nonlinear power spectrum instead of the linear.
The window function for two-dimensional projection for an infinitesimally thin range in radius is simply a Bessel function: $W_{r_p}(k) = 
J_0(kr_p)$.  For a finite range in $r_p$, the window function is the weighted average (by area) over $r_p$:
\beqa
W_{r_p}(k) &=& \frac
{\int_{r_{p,\rm min}}^{r_{p,\rm max}} 2\pi r_p J_0(kr_p)\,\rmd r_p}
{\int_{r_{p,\rm min}}^{r_{p,\rm max}} 2\pi r_p\,\rmd r_p}
\nonumber \\
&=& \frac{2[J_1(kr_{p,\rm max})-J_1(kr_{p,\rm min})]}{k^2(r_{p,\rm max}^2-r_{p,\rm min}^2)}.
\label{eq:wrp}
\eeqa
The constant $C$ accounts for the effect of the integral constraint on the numerator of Eq.~(\ref{eq:lsest}).
There are in principle additional corrections associated with the uncertainty in the denominator (see \citealt{1999ApJ...519..622H} for a thorough 
discussion), which are relevant if the correlation function is of order unity on the scale of the survey region.  This is not the case here so we have 
not included them.

\begin{figure*}
\includegraphics[angle=-90,width=6.5in]{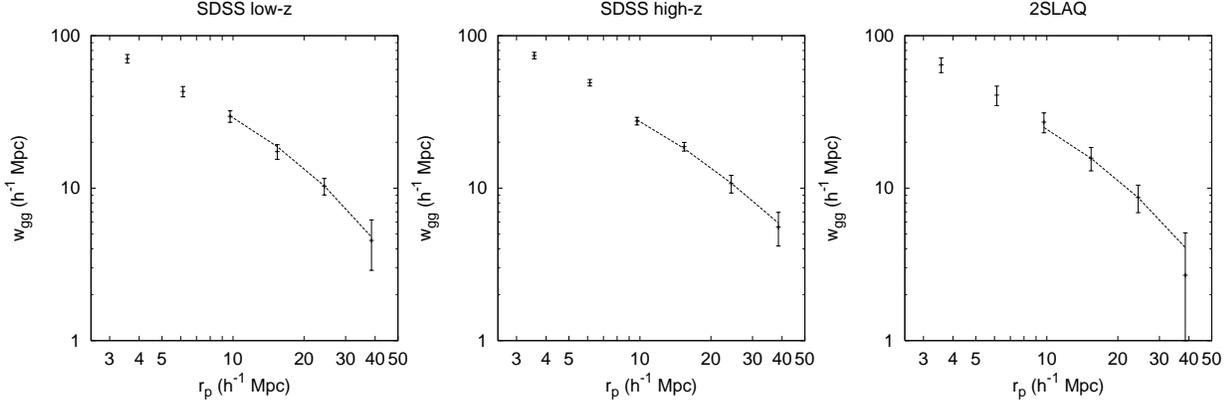}
\caption{\label{fig:bias}The projected galaxy-galaxy correlation functions $w_{gg}(r_p)$ for the three samples and the best-fit nonlinear 
power spectrum (using the \citealt{2003MNRAS.341.1311S} mapping)
to the 7.5--47$h^{-1}\,$Mpc.  Note that the error bars are correlated.}
\end{figure*}

We used the ``WMAP+all'' $\Lambda$CDM cosmology \citep{2006astro.ph..3449S} for the bias determinations, which has $\Omega_m=0.262$, $h=0.708$, 
$\Omega_b=0.0437$, 
$n_s=0.938$, and $\sigma_8=0.751$; in general for other values of $\sigma_8$, the bias scales as $b\propto\sigma_8^{-1}$.  The projected correlation 
functions and best-fit $Q$-models are shown in Figure~\ref{fig:bias}.
The linear theory fit used the \citet{1998ApJ...496..605E} transfer
function; the two 
nonlinear fits differ only in their use of the nonlinear mappings of
either \citet{1996MNRAS.280L..19P} or  
\citet{2003MNRAS.341.1311S}.

The biases of the LSS tracers are displayed in Table~\ref{tab:bias}; note that two values of the minimum $r_p$ were used, with the linear fit 
restricted to the largest scales.  The $b_g$ values are consistent with each other and the $\chi^2$ values are reasonable.  The $Q$ values obtained 
with the $Q$-model fits are $61\pm48$ (SDSS low-$z$), $-24\pm49$ (SDSS high-$z$), and $50\pm 128$ (2SLAQ), i.e. $Q$ is poorly constrained at these 
scales but is consistent with other determinations for similar galaxy
types \citep{2006astro.ph..5302P}.

As a systematics test, we also computed the integral $\int \xi(r_p,\Pi)\,\rmd\Pi$ over the range $60<|\Pi|<120h^{-1}\,$Mpc and computed $b_g^2$ using 
this in place of $w_p(r_p)$.  (For the fit we used the \citealt{2003MNRAS.341.1311S} nonlinear mapping.)  In the absence of systematics, this should 
give a result close to zero since the galaxies considered are well separated along the line of sight, but could be nonzero if (for example) there are 
spurious angular fluctuations in the galaxy distribution.  The procedure gives ``$b_g^2$'' values of $+0.07\pm 0.15$, $-0.06\pm 0.16$, and $+0.17\pm 
0.34$ ($1\sigma$) for the SDSS low-$z$, SDSS high-$z$, and 2SLAQ samples, respectively.  These are all consistent with zero and much less than the 
actual $b_g^2$ values.

\begin{table}
\caption{\label{tab:bias}The bias of the various tracer samples for four fits.
The error estimates are $1\sigma$.  The fitting methods are ``lin'' (linear theory); ``Q'' ($Q$-model); ``PD'' \citep{1996MNRAS.280L..19P}; and ``S'' 
\citep{2003MNRAS.341.1311S}.}
\begin{tabular}{ccccr}
\hline\hline
Sample & $r_p$ range & Method & $b_g$ & $\chi^2$/dof \\
 & $h^{-1}\,$Mpc & & & \\
\hline
$\!$SDSS low-$z\!$
& 12--47  & lin & $1.84\pm 0.15$ & 0.924/1 \\
$\langle z\rangle=0.22$
& 7.5--47 & Q   & $2.05\pm 0.12$ & 0.809/1 \\
& 7.5--47 & PD  & $1.97\pm 0.12$ & 1.476/2 \\
& 7.5--47 & S   & $2.01\pm 0.12$ & 1.428/2 \\
\hline
$\!$SDSS high-$z\!$
& 12--47  & lin & $1.94\pm 0.09$ & 0.001/1 \\
$\langle z\rangle=0.31$
& 7.5--47 & Q   & $1.97\pm 0.08$ & 0.068/1 \\
& 7.5--47 & PD  & $1.93\pm 0.07$ & 1.405/2 \\
& 7.5--47 & S   & $1.97\pm 0.07$ & 1.257/2 \\
\hline
    2SLAQ
& 12--47  & lin & $1.98\pm 0.25$ & 0.879/1 \\
$\langle z\rangle=0.55$
& 7.5--47 & Q   & $2.22\pm 0.24$ & 0.798/1 \\
& 7.5--47 & PD  & $2.10\pm 0.20$ & 1.026/2 \\
& 7.5--47 & S   & $2.13\pm 0.20$ & 0.953/2 \\
\hline\hline
\end{tabular}
\end{table}

For the rest of the paper we have used the biases from the 7.5--47$h^{-1}\,$Mpc fit with the \citet{2003MNRAS.341.1311S} nonlinear power spectrum.  

\subsection{SDSS Main sample}

The density tracer for the SDSS Main sample is the combined L3--6 sample.  The bias of this tracer varies as a function of redshift because the nearby 
part of the sample is dominated by galaxies with $L\le L_\star$ and the more distant part is dominated by galaxies with $L>L_\star$, which show 
stronger clustering.  This should be taken into account when converting $w_{g+}(r_p)$ to $w_{\delta+}(r_p)$; in particular it results in an effective 
bias $b_{\rm eff}$ that varies according to sample whose intrinsic alignments are being measured: the L3 GI correlation function measurement is 
dominated 
by nearby pairs of galaxies, hence $b_{\rm eff}$ is low, while for L6 $b_{\rm eff}$ is high.
A very crude way of estimating this effective bias is to take the pair-weighted average of the bias of the density tracer,
\beq
b_{\rm eff} = \frac{\int b_\delta n_\gamma^{\rm com} n_\delta^{\rm com} \rmd V}{\int n_\gamma^{\rm com} n_\delta^{\rm com} \rmd V},
\label{eq:beff}
\eeq
where $b_\delta$ is the bias of the density tracer, and $n_\gamma^{\rm com}$ and $n_\delta^{\rm com}$ are the comoving number densities of the 
ellipticity
and density tracers, respectively.  For the bias of the density tracers, we use the results from \citet{2004ApJ...606..702T} and 
\citet{2005PhRvD..71d3511S}, namely that $\sigma_8b=0.764$ (L3), 0.848 (L4), 0.968 (L5), and 1.427 (L6).  The resulting values of $b_{eff}$ for each of 
the ellipticity tracers are shown in Table~\ref{tab:main}.  The biases of the L3--6 samples are generally determined to an uncertainty of several per 
cent; we do not explicitly propagate these uncertainties as they will contribute negligibly to the final error in $w_{\delta +}(r_p)$.

\begin{table}
\caption{\label{tab:main}The effective bias $b_{\rm eff}$ of the L3--6 density tracer used in the GI correlation analysis.  As explained in the text, 
$b_{\rm eff}$ depends on the SDSS Main subsample used to trace the ellipticity field because the sample is not volume-limited.
Values are normalized to WMAP ($\sigma_8=0.751$); for other 
values of $\sigma_8$, these values should be re-scaled as $b_{\rm eff}\propto\sigma_8^{-1}$.  The weighted effective redshift $z_{\rm eff}$ at which 
the correlation is measured is defined by replacing $b_\delta$ in Eq.~(\ref{eq:beff}) with $z$.}
\begin{tabular}{ccc}
\hline\hline
Ellipticity tracer & $b_{\rm eff}$ & $z_{\rm eff}$ \\
\hline
L3.red  & 1.08 & 0.07 \\
L3.blue & 1.08 & 0.07 \\
L4.red  & 1.11 & 0.09 \\
L4.blue & 1.12 & 0.09 \\
L5.red  & 1.14 & 0.10 \\
L5.blue & 1.16 & 0.12 \\
L6.red  & 1.16 & 0.12 \\
L6.blue & 1.19 & 0.13 \\
\hline\hline
\end{tabular}
\end{table}

\section{Power law fits for LRGs}
\label{S:fits}

In this section we present an empirical parameterized model for the large-scale GI correlation of LRGs, with power-law dependence on the galaxy's 
luminosity, redshift (technically $1+z$), and transverse separation.  It should be noted that such a model is purely empirical, in the sense that it 
reproduces the correct correlation between the GI amplitude and the specified galaxy properties.  This is all that is needed to predict GI 
contamination for cosmic shear surveys; however when considering theoretical interpretation of the results presented here, it must be remembered that 
our fits represent correlations, not causal relationships.  For example, we find that the GI amplitude scales as roughly $L^{1.5}$, but this does not 
mean that the alignment is ``caused by'' the galaxy's $r$-band luminosity -- in practice both the GI amplitude and the galaxy luminosity are both 
products of the process of galaxy formation and hence are correlated.  In particular, different scalings might be obtained if we also fit dependence on 
e.g. colour, velocity dispersion, surface brightness, etc., since these are correlated with luminosity and redshift.

The simplest model including luminosity, redshift, and scale dependence is
\begin{equation}
w_{\delta +}(r_p) = A_0 \left(\frac{r_p}{r_{\rm pivot}}\right)^\alpha
\left( \frac L{L_0}\right)^\beta \left(\frac{1+z}{1+z_{\rm 
pivot}}\right)^\gamma.
\label{eq:wdp}
\end{equation}
Here there are 4 parameters $\{A_0,\alpha,\beta,\gamma\}$, and $L$ is the 
galaxy luminosity.  We use the $r$-band luminosity $k+e$-corrected to 
$z=0.0$.  The normalization $L_0$ corresponds to absolute magnitude $-22$,
i.e.
\begin{equation}
\frac L{L_0} = 10^{0.4(-22-M_r)}.
\end{equation}
For reference, typical $k+e$-corrections for a completely passive $z=0.3$
LRG are $(1.51, 1.00, 0.15, -0.05, -0.11)$ in the five bands, and for a
$z=0.55$ LRG these are $(2.73, 1.59, 0.77, 0.05, -0.15)$. If the LRG has some
constant low level of star formation, the $k+e$-corrections are $(0.86,
0.92, 0.15, -0.04, -0.10)$ from $z=0.3$ and $(1.04, 1.40, 0.73, 0.05,
-0.13)$ from $z=0.55$.  In practice, $k+e$ corrections are obtained by
considering the observed value of $g-i$ for each galaxy, and doing linear
interpolation using its relationship to the predicted observer-frame $g-i$
values for the completely passive and the passive plus star forming models
at that redshift (at $z=0.3$, the models predict $g-i=2.30$ and $2.14$ for
the two models, respectively; at $z=0.55$, they predict $g-i=2.79$ and
$2.53$).  The pivot points selected here are $z_{\rm pivot}=0.3$ and
$r_{\rm pivot} = 20h^{-1}\,$Mpc.

We have performed a least-squares fit of this $w_{\delta +}(r_p)$ to eight 
LRG samples: the SDSS LRGs split into six samples (the 3 bins in 
luminosity and 2 in redshift), and the 2SLAQ LRGs split into a faint and 
bright sample (separated at $M_r=-22.0$).  We have converted the 
observable $w_{g+}(r_p)$ to $w_{\delta +}(r_p)$ by dividing by the bias of 
the density tracer. For details on the bias determinations, see
Section~\ref{S:bias}.  Note that the bias determination assumed 
$\sigma_8=0.751$; if one wishes to have a constraint on $w_{\delta+}$ for 
other values of $\sigma_8$ it is necessary to multiply the measurement of 
$A_0$ in Table~\ref{tab:lrgfits} by $\sigma_8/0.751$.

The relation $w_{\delta+}(r_p)=w_{g+}(r_p)/b_g$ is the ``obvious'' way to do the conversion from correlation functions involving galaxies to those 
involving matter, but it is worth considering the specific assumptions this makes.  It is valid provided that (i) galaxies are locally linearly biased, 
i.e. the probability distribution for $g({\bmath r})$ conditioned on a particular realization of the density field depends only on $\delta({\bmath r})$ 
and not on $\delta$ at other locations, and that the dependence of the mean galaxy density is linear, $\langle g({\bmath r})\rangle|_\delta =
b_g\delta({\bmath r})$; and (ii) the intrinsic shear of a galaxy depends only on the surrounding density field (or equivalently on the tidal field, 
which contains the same information) and not on the placement of the galaxies, $P({\bmath \gamma}^I|\delta)=P({\bmath 
\gamma}^I|\delta,g)$ (i.e. specifying the positions of other galaxies provides no additional information not in the density field).  Note that the 
dependence of intrinsic shear on the density field is allowed to be nonlocal.  Given these two assumptions we may write, for ${\bmath r}\neq {\bf 0}$,
\beqa
\xi_{g+}({\bmath r}) &=& \langle \tilde{\bmath \gamma}^I_+({\bf 0})g({\bmath r}) \rangle
= \langle [1+\delta({\bf 0})]{\bmath \gamma}^I_+({\bf 0})g({\bmath r}) \rangle
\nonumber \\
&=& \int {\cal D}\delta\; {\cal D}g\; P(\delta,g) \langle {\bmath \gamma}^I_+({\bf 0})\rangle|_{\delta,g}[1+g({\bf 0})]g({\bmath r})
\nonumber \\
&=& \int {\cal D}\delta\; P(\delta)\langle {\bmath \gamma}^I_+({\bf 0})\rangle|_\delta \langle [1+g({\bf 0})]g({\bmath r})\rangle|_\delta
\nonumber \\
&=& \int {\cal D}\delta\; P(\delta)\langle {\bmath \gamma}^I_+({\bf 0})\rangle|_\delta \langle 1+g({\bf 0})\rangle_\delta \langle g({\bmath 
r})\rangle|_\delta
\nonumber \\
&=& \int {\cal D}\delta\; P(\delta)\langle {\bmath \gamma}^I_+({\bf 0})[1+g({\bf 0})]\rangle_\delta b_g\delta({\bmath r})
\nonumber \\
&=& b_g\langle {\bmath \gamma}^I_+({\bf 0})[1+g({\bf 0})]\delta({\bmath r}) \rangle
=b_g\xi_{\delta+}({\bmath r}).
\eeqa
Here the first two lines only involve defintions; the third line uses assumption (ii) to remove the conditioning of $\langle {\bmath \gamma}^I_+({\bf 
0})\rangle$ on $g$; the fourth line makes use of assumption (i) that the galaxy density depends locally on the matter density to split the expectation 
value $\langle [1+g({\bf 0})]g({\bmath r})\rangle|_\delta$; the fifth line uses the linearity of the biasing from assumption (i) to introduce $b_g$, 
and uses assumption (ii) to combine two of the expectation values into $\langle {\bmath \gamma}^I_+({\bf 0})[1+g({\bf 0})]\rangle_\delta$; and the 
last line again only use definitions.
The projected (2-dimensional) relation $w_{\delta+}(r_p)=w_{g+}(r_p)/b_g$ then follows.  (Here $\int {\cal D}\delta$ represents a functional integral 
over realizations of the density field.)  Gaussianity is not assumed, however Gaussianity combined with scale-independent bias (defined by 
$b_g=\sqrt{\xi_{gg}/\xi_{\delta\delta}}$) and unit stochasticity $r_g=1$ would imply the validity of assumption (i) since they completely specify the 
two-point behavior of the galaxy field.  We also note that (i) implies $b_g=$constant and $r_g=1$.  Of our assumptions, (i) is consistent with
(though not uniquely proven by) simulation results suggesting $b_g=$constant and $r_g=1$ down to $\sim 5h^{-1}\,$Mpc \citep{2004ApJ...614..533T}; at 
smaller scales there is a hint of decrease in the bias.  The status of assumption (ii) is unknown, however since galaxy alignments are expected to be 
determined by tidal fields it is physically reasonable.  In the future it would be desirable to do a more detailed analysis, perhaps fitting the 
$w_{g+}(r_p)$ results directly to simulations; for this reason the correlation functions are given in Appendix~\ref{app:data}.  We have also not 
attempted to use the conversion $w_{\delta+}(r_p)=w_{g+}(r_p)/b_g$ at smaller scales than the 4.7--7.5$h^{-1}\,$Mpc bin, although these inner bins are 
provided in the data tables.

We perform several fits to Eq.~(\ref{eq:wdp}).  In each case, since the galaxies used to trace the ellipticity field do not all have the same 
luminosity or redshift, we actually fit
\beq
w_{\delta +}(r_p) = A_0
\frac{\langle r_p^\alpha\rangle}{r_{\rm pivot}^\alpha}
\frac{\langle L^\beta\rangle}{L_0^\beta}
\frac{\langle (1+z)^\gamma\rangle}{(1+z_{\rm pivot})^\gamma},
\eeq
where the luminosity and redshift averages are taken over the galaxies in the sample, and the $r_p$ average is weighted by area (analogously to
Eq.~\ref{eq:wrp}).  In principle one should also weight the redshift average by the comoving number density of the density tracer (full SDSS LRG or 
2SLAQ sample) but in practice the comoving number density is nearly constant for the former and the statistical errors are very large for the latter, 
so we did not implement a correction for this effect.

The fits were done including the correlations between different radial bins for the same subsample, but {\em not} including the correlations between 
different subsamples.  The correlations between different radial bins of $w_{g+}(r_p)$ for the same subsample are clearly seen in any of the covariance 
matrices and are expected because galaxies are clustered.  They must be included to get meaningful results.  However, we do not expect significant 
correlations between $w_{g+}(r_p)$ in different subsamples, so long as the intrinsic alignments are sufficiently weak.  In order to test this, we 
computed the correlation coefficients $\rho_{iA,jB}$ by the jackknife procedure, where we have used the $i,j,...$ indices to denote radial bins and 
$A,B,...$ to denote LRG subsamples.  We consider the 6 radial bins used for fits here ($4.7<r_p<60h^{-1}\,$Mpc) and the 6 SDSS LRG subsamples.  This 
provides 315 cross-correlation coefficients with $A\neq B$.  In each case we then calculate the Fisher $z$-coefficient, defined by $\rho_{iA,jB}=\tanh 
z_{iA,jB}$. These 315 coefficients have a sample mean of $+0.019\pm 0.008$ and a standard deviation of 0.14.  For comparison, with 50 jackknife regions 
of the same size and Gaussian errors, we should have a standard deviation of $\sim 1/\sqrt{50-3}=0.15$, and the mean value of the $\{z_{iA,jB}\}$ 
should be zero if the samples truly are independent.  In fact we observe the correct standard deviation and a hint of correlation between bins at only 
the $\sim 2$ per cent level.  For the two 2SLAQ LRG bins there are 21 $z$-coefficients with a sample mean of $+0.016\pm 0.046$ and a standard deviation 
of 0.21.  Therefore we have not included correlations between different samples.

The fit results are shown in Table~\ref{tab:lrgfits}.  We have included both SDSS-only and SDSS+2SLAQ constraints; it is clear from the fit that the 
SDSS LRG sample provides essentially all of the constraint on $\alpha$ and $\beta$, with the addition of 2SLAQ providing significant new information 
about the redshift evolution parameter, $\gamma$.  This is not surprising since the 2SLAQ sample is small but adds a large baseline in redshift.  Note 
that the errorbars on $A_0$ are rather non-Gaussian, so the apparent $\sim 6\sigma$ effect is actually stronger than that (see the $\Delta\chi^2$ 
values).

The error bars in the table require some explanation.  The usual way of constructing error bars on a single parameter is to change its value until the 
$\chi^2$ (minimized with respect to the other parameters) increases by an amount $\Delta\chi^2=3.84$ (the 95th percentile of the $\chi^2$ distribution 
with 1 degree of freedom).  The problem with this procedure is that if the error bars are determined by resampling (jackknife or bootstrap), the 
resulting covariance matrix ${\mathbfss C}$ of the $w_{g+}(r_p)$ values is noisy (but still unbiased).  This leads to a systematic overestimate of 
${\mathbfss C}^{-1}$ \citep{2004PhRvD..70f3526H, 2006MNRAS.367..611M, 2006astro.ph..8064H}.  In particular, if one has a set of parameters $\{p^i\}$ 
being fit, the shift $\Delta\chi^2_i$ (i.e. the $\chi^2$ using the true value of parameter $i$ minimized over the other parameters, minus the $\chi^2$ 
minimized over all parameters, and using the estimated covariance matrix) is not distributed as a $\chi^2$ with 1 degree of freedom, even for 
Gaussian-distributed errors.  A solution to this problem was proposed by \citet{2004PhRvD..70f3526H} and \citet{2006MNRAS.367..611M} who showed that if 
the data have Gaussian errors, the covariance matrix is obtained from Gaussian data, and the model is linear in the parameters, then the distribution 
of $\Delta\chi^2_i$ does not depend on the covariance matrix ${\mathbfss C}$ and hence can be obtained from a Monte Carlo simulation.  Unfortunately 
their procedure does not apply directly to our problem because we only estimated the covariance matrix entries between $w_{g+}(r_p)$ values for the 
same subsample -- the cross-entries between different subsamples are zero.  We have circumvented this problem by doing a Monte Carlo simulation as 
follows.  First, we take the jackknife covariance matrix ${\mathbfss C}$ to be the fiducial model and simulate the re-sampling by the procedure 
described in Appendix D of \citet{2004PhRvD..70f3526H} with $M=50$ regions to get a re-sampled matrix $\hat{\mathbfss C}$.  We then set the 
cross-entries between different subsamples to zero.  We also generate a simulated set of estimated $\{\hat w_{g+}(r_p)\}$ according to a Gaussian 
distribution with mean given by the best-fit power law model and covariance ${\mathbfss C}$.  From this simulated data and simulated covariance matrix, 
one can then compute a $\chi^2$ surface and find $\Delta\chi^2_i$ for each of the four parameters $\{p^i\}_{i=0}^3$.  Repeating this procedure 1024 
times allows us to compute the $\Delta\chi^2_i$ distribution for each parameter.  The 95th percentile of this distribution is shown in 
Table~\ref{tab:lrgfits} for each fitting region and each parameter, and is used to compute the 95 per cent confidence limits on the parameters.

\begin{table*}
\caption{\label{tab:lrgfits}The best-fit parameters to 
Eq.~(\ref{eq:wdp}) using SDSS and 2SLAQ LRGs.  Error bars are 95 per cent confidence limits.
The first set of fit parameters uses the most conservative cut on transverse separation.  The $\Delta\chi^2$ value in the 
second-to-last column is the improvement in $\chi^2$ relative to no GI correlation, and the 
$\Delta$dof value indicates the number of parameters in the fit.
The amplitude $A_0$ is given in units of 0.01$h^{-1}\,$Mpc.  The $\Delta\chi^2_i$ values in the last column indicate the degradation of $\chi^2$ used 
to compute the 95 per cent confidence limits on $A_0$, $\alpha$, $\beta$, and $\gamma$, respectively.  These are greater than 3.84 for 1 parameter 
because of noise in the covariance matrix (see text).}
\begin{tabular}{cccccccc}
\hline\hline
Fit region & $A_0/(0.01h^{-1}\,$Mpc$)$ & $\alpha$ & $\beta$ & $\gamma$ &
$\chi^2$/dof
& $\Delta\chi^2/\Delta$dof & $\Delta\chi^2_i$ \\
\hline
\multicolumn{8}{c}{Fits to SDSS+2SLAQ}\\
\hline
$r_p>11.9h^{-1}\,$Mpc &
$+6.0^{+2.6}_{-2.2}$ &
$-0.88^{+0.31}_{-0.34}$ &
$+1.51^{+0.73}_{-0.69}$ &
$-1.00^{+2.40}_{-3.19}$ &
33.3/28 & 171.8/4 & 4.38, 4.74, 4.16, 4.47 \\
$r_p>7.5h^{-1}\,$Mpc &
$+6.4^{+2.5}_{-2.1}$ &
$-0.85^{+0.24}_{-0.25}$ &
$+1.41^{+0.66}_{-0.63}$ &
$-0.27^{+1.88}_{-2.46}$ &
42.8/36 & 215.8/4 & 5.10, 4.89, 5.07, 5.02 \\
$r_p>4.7h^{-1}\,$Mpc &
$+5.9^{+2.3}_{-2.0}$ &
$-0.73^{+0.19}_{-0.19}$ &
$+1.48^{+0.64}_{-0.63}$ &
$-0.56^{+2.02}_{-2.74}$ &
54.9/44 & 219.2/4 & 5.04, 5.08, 4.92, 5.57 \\
\hline
\multicolumn{8}{c}{Fits to SDSS only}\\
\hline
$r_p>11.9h^{-1}\,$Mpc &
$+7.1^{+3.4}_{-2.7}$ &
$-0.95^{+0.32}_{-0.35}$ &
$+1.43^{+0.73}_{-0.71}$ &
$+1.94^{+4.75}_{-4.52}$ &
21.3/20 & 173.9/4 & 4.49, 4.67, 4.27, 4.27 \\
$r_p>7.5h^{-1}\,$Mpc &
$+7.4^{+2.9}_{-2.4}$ &
$-0.88^{+0.24}_{-0.25}$ &
$+1.31^{+0.67}_{-0.66}$ &
$+2.39^{+4.52}_{-4.30}$ &
27.9/26 & 208.7/4 & 4.66, 5.13, 4.91, 4.66 \\
$r_p>4.7h^{-1}\,$Mpc &
$+6.6^{+2.7}_{-2.2}$ &
$-0.74^{+0.19}_{-0.18}$ &
$+1.44^{+0.63}_{-0.62}$ &
$+1.81^{+4.52}_{-4.40}$ &
34.0/32 & 213.3/4 & 4.88, 5.10, 4.66, 5.00 \\
\hline\hline
\end{tabular}
\end{table*}

The question naturally arises whether the fits to the LRG signal in
Table~\ref{tab:lrgfits} apply to the GI correlation for the
red galaxies fainter than the LRG1 and 2SLAQ-faint samples.    Certainly in making a model for the GI 
correlations
for cosmic shear purposes it is necessary to have a model for these
fainter objects, which are  
after all much more numerous.  The best way to assess this is to take
our LRG models and compare them with the measurements for the L3, L4,
L5, and L6  
samples of red Main galaxies.  It is seen from
Figure~\ref{F:allcolorconf} that our best-fit value of $\alpha=-0.73$ is
consistent with the slope in all  
four cases, so it remains to test the amplitude.  We do this by taking
the GI measurements for each of these samples at $r_p>4.7h^{-1}\,$Mpc
and fitting to them a power-law
\beq
w_{\delta+}(r_p) =w_{\delta+}(r_{\rm pivot}) \left(\frac{r_p}{r_{\rm pivot}}\right)^\alpha,
\eeq
where $\alpha$ is fixed to $-0.73$ and the normalization
$w_{\delta+}(r_{\rm pivot})$ is allowed to vary to minimize the
$\chi^2$.  The resulting  
amplitudes are shown in Table~\ref{tab:red}.  The L5 and L6 samples
actually include some of the SDSS LRGs, so only the comparisons for L3
and L4  
represent an independent test of the power-law model.  There are
detections at $>2\sigma$ for the L3, L5, and L6 red samples, although
the  
significance is far greater for L6 than for the others.  We have also
displayed, in the last column, the amplitude predicted by the best LRG
fit, after converting the Petrosian magnitudes measured for the SDSS Main galaxies to model magnitudes by dividing by 0.8 (the fraction of the flux 
captured by the Petrosian method for a typical elliptical galaxy; \citealt{2001AJ....121.2358B}).  This  
amplitude is consistent with all of the samples at the 95 per cent
confidence level. 

\begin{table}
\caption{\label{tab:red}The GI correlation amplitude measured from the L3--L6 red Main samples of galaxies, and that predicted by the best LRG fit 
(to SDSS+2SLAQ, $r_p>4.7h^{-1}\,$Mpc data).  The pivot radius is $r_{\rm pivot}=20h^{-1}\,$Mpc.  Note that the L5 and L6 samples have some overlaps 
with the LRG sample, so that for these samples the two columns are not independent.  Errors shown are 95 per cent confidence limits, which for 50 
jackknife regions, 6 radial bins, and 1 parameter (the amplitude) being fit corresponds to $\Delta\chi^2=5.02$.}
\begin{tabular}{ccc}
\hline\hline
Sample & \multicolumn{2}{c}{$w_{\delta+}(r_{\rm pivot})$ ($h^{-1}\,$Mpc)} \\
 & Measured (95\%CL) & Predicted from LRGs \\
\hline
L3.red & $+0.035\pm 0.032$ & $+0.004$ \\
L4.red & $+0.013\pm 0.019$ & $+0.013$ \\
L5.red & $+0.024\pm 0.022$ & $+0.044$ \\
L6.red & $+0.144\pm 0.078$ & $+0.132$ \\
\hline\hline
\end{tabular}
\end{table}

\section{Comparison to theory and simulations}
\label{S:theory}

In this section we compare our results to analytical and simulation-based models of intrinsic alignments.

\subsection{Analytical models}

It is worth comparing the results of this analysis to theoretical expectations. \citet{2004PhRvD..70f3526H} presented an analytic ``linear passive'' 
model for red galaxies, which assumes that a galaxy's ellipticity is a linear function of the local tidal quadrupole when the galaxy forms and then 
remains unchanged.  In this case we expect that on linear scales we should have $P_{\delta,\tilde\gamma^I}(k,z)\propto k^{n_s^{\rm eff}}D(z)$, where 
$n_s^{\rm eff}$ is the scalar spectral index corrected for the transfer function, i.e. $n_s^{\rm eff}=n_s+2\rmd[\ln T(k)]/\rmd\ln k$, and $D(z)$ is the 
growth factor.  The redshift dependence is simply the growth factor because $\delta$ increases but $\tilde\gamma^I$ does not, and the scale dependence 
is simply that of the underlying power spectrum.  [This should be equivalent to Eq.~(18) of \citet{2004PhRvD..70f3526H}, since there 
$\bar\rho\propto(1+z)^3$, $\bar D(z)\equiv (1+z)D(z)$, and $P_\delta^{\rm lin}(k,z)\propto k^{n_s^{\rm eff}}D^2(z)$.  There is still a factor of 
$(1+z)^2$ that is missing from Eq.~(14--18) of \citet{2004PhRvD..70f3526H} because of the conversion from comoving to physical scale in the 
potential-density relation.]  Transforming to real space, this predicts:
\beq
\alpha = -2-n_s^{\rm eff} \quad{\rm and}\quad
\gamma = \frac{\rmd\ln D(z)}{\rmd\ln(1+z)}.
\eeq
[There is a $2$ instead of a $3$ in $\alpha$ because $w_{\delta+}(r_p)$ is 
a projected quantity and hence is obtained from 
$P_{\delta,\tilde\gamma^I}(k)$ by a two-dimensional Fourier transform.]  
For a $\Lambda$CDM cosmology, this predicts that $\gamma$ should rise from 
$-1$ at high redshift (matter domination) to $0$ in the far future 
($\Lambda$ domination).  Across the range of redshifts considered here, 
0.2--0.7, we expect $\gamma\sim -0.7$.
The prediction for $\alpha$ depends somewhat on scale since $n_s^{\rm 
eff}$ is not constant.  We can find the value of $\alpha$ relevant to our 
observations by taking $P_{\delta,\tilde\gamma^I}(k)\propto P_\delta^{\rm linear}(k)$ 
and using the Hankel transform
\beq
w_{\delta+}(r_p) = -\frac 1{2\pi}\int P_{\delta,\tilde\gamma^I}(k)
J_2(kr_p)k\,dk
\label{eq:w-p}
\eeq
to get $w_{\delta+}(r_p)$.  Measured across the range from 
(11.9--60)$h^{-1}\,$Mpc (i.e. using the largest-scale cutoff from 
Table~\ref{tab:lrgfits}), we find $\alpha=-0.65$ for the fiducial 
cosmology.  Including smaller scales leads to an increase in $\alpha$ 
because the power spectrum curves downward, $\rmd^2[\ln P(k)]/\rmd(\ln 
k)^2<0$: we have $\alpha=-0.52$ over the range (7.5--60)$h^{-1}\,$Mpc, and 
$\alpha=-0.41$ over the range (4.7--60)$h^{-1}\,$Mpc.  The
linear model does not give a prediction for $\beta$, which requires an 
understanding of how the proportionality constant between ellipticity 
and tidal field relates to the luminosity.

Our results for red galaxies are entirely consistent with the prediction
$\gamma=\gamma_{\rm passive}\sim-0.7$, although given the large error bars
in Table~\ref{tab:lrgfits} this is not a particularly impressive
accomplishment.  From a practical perspective,
we have at least set an upper limit on $\gamma$, which enables upper
bounds on contamination to be placed in cosmic shear investigations.  
From a theoretical perspective, we are still unable to answer perhaps the
most interesting question: is $\gamma$ larger or smaller than the passive
evolution prediction?  A value of $\gamma>\gamma_{\rm passive}$ would
imply that the intrinsic alignments of red galaxies were greater in the 
past, and were being reduced, perhaps due to relaxation, merger, 
or figure rotation processes that destroy pre-existing correlations.  
Conversely, a value of $\gamma<\gamma_{\rm passive}$ would imply that some 
process was causing the LRGs to align themselves with the large-scale 
density field, even at low redshift where LRGs are generally believed to 
be passively evolving.

By contrast, the linear theory predictions for $\alpha$ are only in marginal agreement with observations, with the data giving a smaller (more 
negative) $\alpha$.  The discrepancy is not significant if one uses the range (11.9--60)$h^{-1}\,$Mpc; however one finds a 
discrepancy of $\Delta\chi^2=9.02$ ($p=0.005$ using the distribution from the Monte Carlo simulations) if one includes data down to $7.5h^{-1}\,$Mpc or 
$\Delta\chi^2=13.11$ ($p<10^{-3}$) down to $4.7h^{-1}\,$Mpc.  The direction of this disagreement 
is what one would expect if nonlinear clustering on small scales enhanced $w_{\delta+}(r_p)$, since this would tilt $\alpha$ to more negative values 
than predicted by linear theory.  Indeed if one substitutes the nonlinear power spectrum $P_\delta^{\rm nl}(k)$ of \citet{2003MNRAS.341.1311S} into 
Eq.~(\ref{eq:w-p}) then the predicted values of $\alpha$ are $-0.72$, $-0.68$, and $-0.67$ for $r_{p,\rm min}=11.9$, 7.5, and $4.7h^{-1}\,$Mpc 
respectively.  These are in very good agreement ($\le 1\sigma$) with the measured slopes, however the theoretical justification for believing 
$P_{\delta,\tilde\gamma^I}(k)$ to trace the matter power spectrum in the nonlinear regime is dubious.

The data on GI correlation for blue galaxies are too noisy to constrain their shape, so we have not attempted to compare this to theoretical models.  
We also note that the main theory for alignments of these galaxies,
namely the tidal torque theory
\citep{1969ApJ...155..393P,1970Ap......6..320D,1984ApJ...286...38W},
generically predicts zero GI correlation  
up to second order in perturbation theory \citep{2002astro.ph..5512H, 2004PhRvD..70f3526H}.  Thus more detailed theoretical calculations will be 
necessary to predict the scale dependence of $w_{\delta+}(r_p)$ for blue galaxies.

\subsection{Simulations}
\label{ss:comparison}

The GI contamination to the lensing signal has been investigated in simulations by \citet{2006MNRAS.371..750H}. They used $\Lambda$CDM $N$-body 
simulations populated using the \citet{2005ApJ...627L..89C} conditional luminosity function and several different models for determining the galaxy (as 
opposed to halo) ellipticity.  They then computed the GI correlations using various models: an ``elliptical'' model for which galaxies were assigned 
the same ellipticity magnitude and direction as the parent halos; a ``spiral'' model for which they were considered to be disks with some random 
misalignment with the parent halo angular momentum vector; and a ``mix'' model with a mixture of the above.  They concluded that only the mix model was 
consistent with the measurements from \citet{2006MNRAS.367..611M}.

\begin{figure}
\includegraphics[angle=-90,width=3.2in]{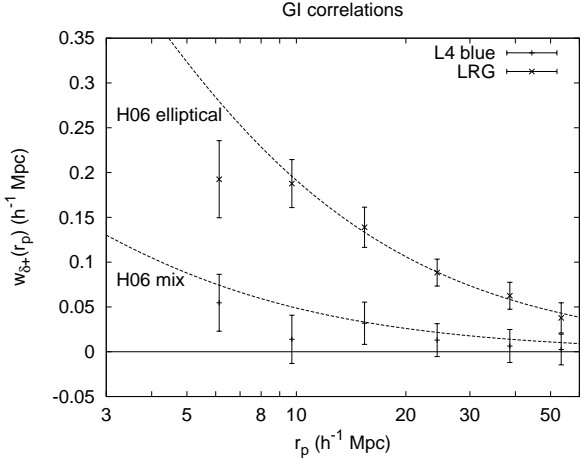}
\caption{\label{fig:h06}The GI correlation functions $w_{\delta+}(r_p)$ for the full SDSS LRG sample and the L4 blue sample, with the galaxy-to-density 
conversion using biases of 1.99 and 1.12 respectively.  The lines are the curves from \citet{2006MNRAS.371..750H} at $z=0.27$, interpreted using the 
procedure in Appendix~\ref{app:heymans}.  The median redshift of the LRGs is $z_{\rm med}=0.27$; the L4 blue galaxies have $z_{\rm eff}=0.12$, 
although we expect that evolution between 0.12 and 0.27 would make a small effect.}
\end{figure}

Here it is possible to do a more detailed comparison.  In Appendix~\ref{app:heymans}, we derive the conversion from the intrinsic shear-lensing shear 
correlation functions measured by \citet{2006MNRAS.371..750H} to $w_{\delta+}(r_p)$.  We have applied this conversion to the elliptical and mix models, 
and plotted these in Figure~\ref{fig:h06}.  Overplotted are the $w_{\delta+}(r_p)$ data for the LRGs (median magnitude $\sim 2$ magnitudes brighter 
than $L_\star$) and the L4 blue galaxies (i.e. $\sim L_\star$).  The LRG data are consistent with the \citet{2006MNRAS.371..750H} ``elliptical'' model, 
which is the one with the strongest GI correlation.  This may seem surprising since the ``elliptical'' model was designed to be a maximal estimate of 
GI, in the sense that the galaxies were assumed to trace the ellipticity of their host haloes perfectly -- one would expect that in the real Universe 
there would be some misalignment.  However the ``elliptical'' model is also averaged over a range of halo masses.  Since LRGs typically occupy the most 
massive haloes, and simulations have suggested that the density-halo ellipticity correlation is stronger at high masses \citep{2005ApJ...618....1H}, it 
is possible that \citet{2006MNRAS.371..750H} would have found a much stronger GI correlation if they had only considered LRGs.  Simulation results 
broken into luminosity bins should thus be a priority: until they are available we cannot say whether our LRG measurements confirm the 
\citet{2006MNRAS.371..750H} ``elliptical'' model, or if there is substantial misalignment in the real Universe that fortuitously results in agreement 
with \citet{2006MNRAS.371..750H} because of their lower typical halo mass.  In any case we note that LRGs show the strongest alignment signal of all 
galaxy types considered and make up only a small fraction of any flux-limited sample of galaxies, so for a realistic weak lensing survey the GI 
contamination would be lower.  The more abundant $L_\star$ blue galaxies are consistent (within $2\sigma$) of either zero alignment or the 
\citet{2006MNRAS.371..750H} ``mix'' model, and it is evident that more data will be needed to distinguish these possibilities.

We are not aware of any simulation results that explore the luminosity dependence of the GI correlation.  Studies of {\em cluster} alignments in 
$N$-body simulations \citep{2005ApJ...618....1H} find a preferential alignment of the major axis of the cluster with the direction to neighbouring 
clusters with a strength that increases with halo mass.  This is qualitatively consistent with what we observe for the alignments of LRGs, which 
increase with luminosity, but a quantitative comparison would require one to populate the haloes with galaxies, assign ellipticities, and convert 
alignment angles to density-shape correlation functions; each of these (but especially the latter two) is a significant source of uncertainty.

\section{Estimates of contamination}\label{S:contam}

In principle, with knowledge of the GI correlation scaling with luminosity, colour, redshift, and transverse separation, combined with knowledge of the 
joint luminosity-colour-redshift distribution for a particular survey, we can predict the contribution of GI contamination to the measured cosmic shear 
power spectrum.  In principle, this contamination then carries over to an underestimate of $\sigma_8$ or, if the evolution of the amplitude is measured 
to constrain the equation of state of dark energy, there will be errors in the measurement of $w_0$ and $w_a$.  In future work, we will attempt to 
quantify more precisely the effects on the dark energy parameter estimates. For now, we merely present the fractional contamination of the cosmic shear 
power spectrum measurement for model surveys with realistic redshift and luminosity distributions, and present a prescription for marginalizing over GI 
uncertainties in measurements of $\sigma_8$.

\subsection{Contamination models}

To make a prediction of contamination to cosmic shear results, we start with a flux limit, and must assume our model fits for intrinsic alignment 
amplitude as a function of 
separation, redshift, and luminosity; a luminosity function as a function of spectral type; the underlying cosmology (to get the distance modulus); and 
$k+e$-corrections.  From this, we can predict the redshift distribution and the magnitude of the GI correlation contribution to the measured cosmic 
shear power spectrum.  In principle, there is also a correction associated with the types of magnitudes used, e.g. Petrosian vs. model.  For example, 
the Petrosian system typically misses $\sim 20$ per cent of the flux
for well-resolved elliptical galaxies \citep{2001AJ....121.2358B}.  Since the GI signal for red galaxies scales as roughly 
$L^{1.5}$, this would translate into roughly a 30 per cent error in the GI amplitude, which is small compared to the redshift extrapolation uncertainty 
in our models.  To the extent that it matters, users of these models should note that our contamination estimate for red galaxies is fit to the 
LRGs, for which we used model magnitudes; for blue 
galaxies the constraints on GI are so weak that changes of this order are unimportant.

For the $r$-band luminosity function as a function of spectral type
(including luminosity evolution), and for
$k$-corrections, we rely on the results from COMBO-17 for
$0.2<z<1.2$ \citep{2003A&A...401...73W}.  For a typical cosmic shear survey that is dominated by 
galaxies with $0.8<z<1.2$, the galaxies that are most important for
the intrinsic alignments are those at the lower end of the redshift
range, where our assumptions about luminosity evolution, etc. are most
likely to be valid.  For the purposes of this work, the spectral types
1 and 2 in that paper are considered ``red,'' types 3 and 4 are
considered ``blue.''  The reason for this distinction is that at
$z=0$, integration of the templates to obtain observed AB $u-r$ colours
(which were used in this paper for colour separation) suggests that
all type 1 and nearly all type 2 galaxies would have been classified
as red, whereas types 3 and 4 would have been classified as blue.  At
redshift 0.1--0.25, the classification of the type 2 template as red
versus blue is no longer clear, but this uncertainty is in part due to
certain features of the spectrum below $300$ nm redshifting into
the $u$-band.  These features have changed significantly in updated
versions of the spectra used by the COMBO-17 team, since the templates
of \cite{2003A&A...401...73W} were found to be inadequate for full
galaxy classification\footnote{Christian Wolf, private communication.},
so we define our correspondance between SDSS and COMBO-17 types using the $z=0$
colour (which is not susceptible to influence by these
uncertainties in the $\lambda<300\,$nm portion of the spectrum).  With this
classification scheme, roughly $25$ per cent of the galaxies in a
flux-limited survey to $R=24$ are classified as red.

The intrinsic alignment signal for the red galaxies is very well constrained at $z\sim 0.3$, with the major uncertainty being the scaling of the 
signal with redshift.  The situation is very different for the blue galaxies for which we have no detection, and for which we only have a
low-redshift constraint: the GI correlation for these objects 
may be near present upper limits, or alternatively could be negligible.  In light of these uncertainties, we have defined four different models, all 
based on the fits at $r_p>4.7h^{-1}\,$Mpc:
\newcounter{models}
\begin{list}{\arabic{models}. }{\usecounter{models}}
\item In the ``pessimistic'' model (A), we assume that the redshift scaling of GI for the red galaxies is the 95 per cent confidence upper limit from 
the SDSS+2SLAQ fit with $r_p>4.7$ \hmpc, $\gamma=+1.47$.  The GI values for the blue galaxies are taken to have the same radial scaling
$\alpha$ and redshift scaling $\gamma$ as for the red galaxies, with the amplitude taken to be the 95 per cent confidence upper limit from
SDSS Main.  The use of the same radial scaling $\alpha$ is motivated by the expectation that the alignment on large scales should trace the tidal
field, which has the same spatial dependence for all types of galaxies.  A single amplitude is fit to the L3--5 blue galaxies to improve statistics 
since there is no sign of a detection in any of these bins.  A separate amplitude is used for the L6 blue galaxies as these have a marginal 
detection when $\alpha$ is fixed.
\item In the ``central'' model (B), we use the best-fit values of amplitudes 
and power law indices $\alpha,\beta,\gamma$ from Table~\ref{tab:lrgfits} using both SDSS and 2SLAQ and including radii down to 4.7$h^{-1}\,$Mpc.  For 
the blue galaxies we use the same value of $\alpha$, as explained above, but fix $\gamma$ to the passive value, $\gamma_{\rm passive}=-0.7$ and use the 
best-fit rather than worst-case value for the amplitude.  The L3--5 galaxies are combined, just as for model B.
\item In the ``optimistic'' model (C), we treat the red galaxies and the L6 blue galaxies just 
as for model B.  We assume the L3--5 blue galaxies have no GI signal, since none is required by the data.
(We do not view the detection in the L6 blue galaxies as robust since the significance is $<2.5\sigma$ and only appears
in one of the four blue bins.)
\item In the ``very optimistic'' model (D), we take the minimum value of $\gamma$ allowed by SDSS+2SLAQ at the $2\sigma$ level ($\gamma=-2.81$), and
use the values of the other parameters that minimize the $\chi^2$ constrained to fixed $\gamma$: $A_0=0.056$, $\alpha=-0.73$, and $\beta=1.44$.  Just 
as for model C, we assume no alignment for L3--5 blue galaxies.  We take the $2\sigma$ lower limit for L6 blue galaxies.
\end{list}
Note that in each model, we used constrained (i.e. $\alpha$ and $\gamma$ fixed) fits to the blue galaxies in each of the luminosity bins in the 
$r_p=(4.7$--$60)h^{-1}\,$Mpc regime.  The best-fit amplitudes $A=w_{\delta+}(20h^{-1}\,$Mpc$)$ assuming the model A scale dependence ($\alpha=-0.71$) 
are $+0.003\pm 0.026$, $+0.017\pm 0.022$, $-0.015\pm 0.035$, and $+0.27\pm 0.25$ $h^{-1}\,$Mpc for the L3, L4, L5, and L6 blue samples respectively, 
where 95 per cent confidence errors are given.  For model B/C/D scale dependence ($\alpha=-0.73$), we find very similar results: 
$w_{\delta+}(20h^{-1}\,$Mpc$) = +0.003\pm 0.026$, $+0.017\pm 0.022$, $-0.015\pm 0.034$, and $+0.27\pm 0.25$ $h^{-1}\,$Mpc for the same samples.  For 
presentation (Table~\ref{tab:models}) we have re-scaled all values to $z=0.3$ using the specified choice of $\gamma$ and the mean redshifts of the 
samples (see Table~\ref{tab:main}).  In each case we have used the ``L3--5 blue'' amplitude for the blue galaxies fainter than L3, since we have no 
better constraint for them (the SDSS Main sample does not probe an interesting volume at fainter luminosities).

The models are summarized in Table~\ref{tab:models}, along with estimates of the fractional GI contamination as quantified by the ratio 
$C_l^{GI}/C_l^{GG}$.  For most cosmic shear studies, model A can be taken as a ``$2\sigma$'' estimate of the GI 
contamination.  Model B should be viewed as a ``best guess,'' with the caveat that we have no detection of GI in the L3--5 blue samples.  If the reason 
for this nondetection is that the GI signal in these samples is {\em much} less than the current upper limits, then model C may be more realistic than 
B.  

\begin{table*}
\caption{\label{tab:models}The four GI models used here to assess contamination.  The power law indices are defined by
$w_{\delta +}(r_p)\propto r_p^\alpha(1+z)^\gamma$.  The amplitudes are normalized to $\sigma_8=0.751$; for cosmologies with different values of 
$\sigma_8$, the bias of the density tracer scales as $b\propto\sigma_8^{-1}$ and the GI contamination scales as $w_{\delta+}\propto\sigma_8$.  
``Fractional contamination'' here is defined by $C_l^{GI}/C_l^{GG}$.}
\begin{tabular}{clllllcc}
\hline\hline
\multicolumn{2}{c}{Model} & Galaxy & \multicolumn{2}{c}{Power law indices} & $w_{\delta+}(20h^{-1}\,$Mpc$,z=0.3)$ &
\multicolumn{2}{c}{Fractional contamination} \\
 & & type & $\alpha$ & $\gamma$ & $h^{-1}\,$Mpc & $R<23$, $l=500$ & $R<24$, $l=500$ \\
\hline
A & Pessimistic & red & $-0.71$ & $+1.47$ & $+0.056(L/L_0)^{1.58}$  & $-0.420$ & $-0.332$ \\
  &             & blue& $-0.71$ & $+1.47$ & $+0.028$ (L3--5); $+0.61$ (L6) & & \\
\hline
B & Central     & red & $-0.73$ & $-0.56$ & $+0.059(L/L_0)^{1.48}$ & $-0.103$ & $-0.065$ \\
  &             & blue& $-0.73$ & $-0.70$ & $+0.005$ (L3--5); $+0.25$ (L6) & & \\
\hline
C & Optimistic  & red & $-0.73$ & $-0.56$ & $+0.059(L/L_0)^{1.48}$ & $-0.042$ & $-0.021$ \\
  &             & blue& $-0.73$ & $-0.70$ & 0 (L3--5); $+0.25$ (L6) & & \\
\hline
D & Very
    optimistic  & red & $-0.73$ & $-3.29$ & $+0.055(L/L_0)^{1.43}$ & $-0.034$ & $-0.015$ \\
  &             & blue& $-0.73$ & $-3.29$ & 0 (L3--5); $+0.03$ (L6) & & \\
\hline\hline
\end{tabular}
\end{table*}

\begin{figure}
\includegraphics[angle=-90,width=3.2in]{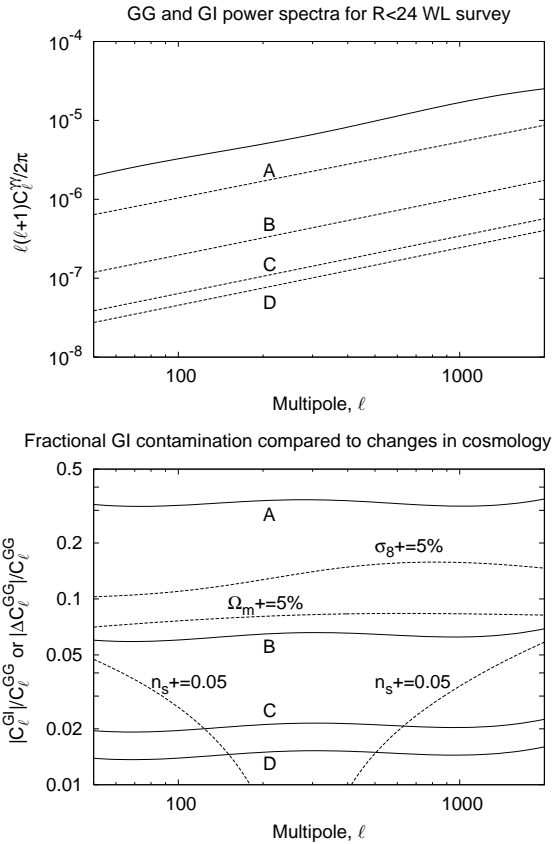}
\caption{\label{fig:ref-contam}{\em Upper panel:} The cosmic shear power spectrum (solid line) and GI power spectra (dashed lines) for all galaxies 
with $R<24$.  We plot the absolute value of GI since it is negative for all models.
{\em Lower panel:} The fractional contamination of the power spectrum $|C_l^{GI}|/C_l^{GG}$ (solid lines), compared to the change produced by several 
changes in the cosmological parameters $|\Delta C_l^{GG}|/C_l^{GG}$ (dashed lines).  This 
ranges from $\sim 33$ per cent for model A to $\sim 1.5$ per cent for model D.  We have plotted the range out to $l=2000$ ($k\sim 1h\,$Mpc$^{-1}$ at 
the typical source redshift) since at smaller scales we cannot convert from $w_{g+}(r_p)$ to $w_{\delta +}(r_p)$ (see Section~\ref{S:fits}).  For the 
variations of cosmological parameters, $\Omega_m$ is varied at fixed $\Omega_mh^2$ and $n_s$ is varied at fixed $\sigma_8$.}
\end{figure}

As an example of these models, we show in Figure~\ref{fig:ref-contam} the GI contamination for a cosmic shear survey that measures galaxies down to 
$R=24$ (median redshift 0.6).  The results in the figure are not necessarily representative of cosmic shear surveys to the specified depth, since the 
sources will not all be weighted equally (and some may be rejected from the analysis due to being poorly resolved, or having poorly constrained 
photo-$z$s with the particular bandpasses used in the survey). The fractional contamination $C_l^{GI}/C_l^{GG}$ is $-33$, $-6.5$, $-2.1$, and $-1.5$ 
per cent for models A--D respectively at $l=500$ (these values are given in Table~\ref{tab:models}).  As can be seen from the figure, these fractions 
are 
almost independent of $l$ because the intrinsic 
alignment power spectrum $P_{\delta,\tilde\gamma^I}(k)\propto k^{-2-\alpha} = k^{-1.27\pm 0.19}$ (95\%CL) has roughly the same $k$-dependence as the 
matter power spectrum.  This was actually measured for the LRGs; setting the blue galaxies (which have no detection) to the same slope as the red 
galaxies was a modeling assumption.  The fraction of the contamination coming from blue galaxies is 93, 69, 3.3, and 0.4 per cent for models A--D.  It 
is clear that refining the models for blue galaxies should be a high priority for future work since these dominate the uncertainty in the GI estimates 
(though it is not clear whether they dominate the GI contamination).  Note that the contamination estimates presented are on the power spectrum; the 
fractional contamination on the amplitude $\sigma_8$ will be less by a factor of $\sim 2$ since the power spectrum is roughly proportional to 
$\sigma_8^2$.  In particular for the survey to $R=24$ at $l=500$, the reduction in shear power spectrum $C_l^{\gamma\gamma}$ is equivalent to changing 
$\sigma_8$ by $\Delta\sigma_8=-0.10$, $-0.02$, $-0.005$, and $-0.004$ for models A--D, respectively.

Integrated over all scales, the GI contamination we predict in the pessimistic model (A) results in a 1$\sigma$ error for a survey measuring all 
galaxies to $R=24$ with combined shape and measurement noise $\gamma_{\rm rms}=0.3$ over a region of sky of only $4\,$deg$^2$.  Thus if model A is 
correct then GI contamination may already be important for the current generation of lensing surveys.  Models B, C, and D predict an error of 1$\sigma$ 
for sky coverage of 100, 1000, and 2000$\,$deg$^2$ respectively; thus even in the optimistic cases GI contamination will be significant and will have 
to be removed in future surveys.

\begin{figure}
\includegraphics[angle=-90,width=3.2in]{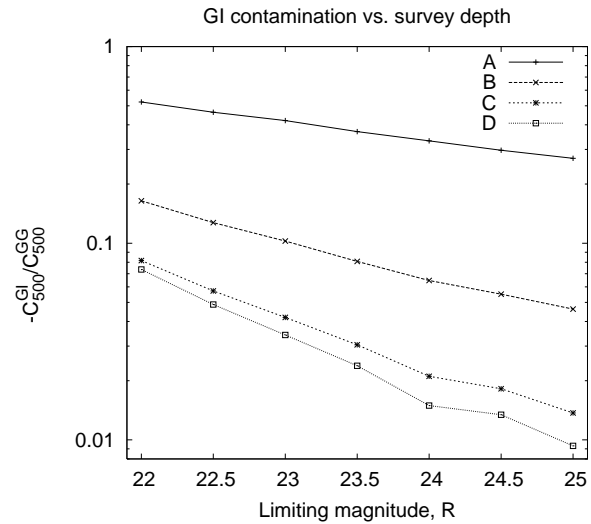}
\caption{\label{fig:rmag}The GI contamination at $l=500$ for the four models as a function of survey depth.  We considered all galaxies with $R$-band 
magnitude less than the specified cutoff, and $z<1.2$.}
\end{figure}

We have also investigated the dependence of the contamination estimates on the magnitude cut, as shown in Figure~\ref{fig:rmag}.  For this calculation 
the COMBO-17 luminosity function was also used, hence a cutoff in the redshift distribution at $z=1.2$ was imposed.  The amount of contamination 
generally decreases with survey depth: for our ``central'' model it is twice as severe at $R_{\rm max}=22.5$ than at $R_{\rm max}=24$.

We also could have constructed a more extreme pessimistic model than model A by taking the 95 per cent confidence upper limit in each of bins L3, 
L4, and L5 separately; for the above-mentioned toy cosmic shear survey to $R=24$, this ``AA'' model leads to $-41$ per cent contamination at $l=500$ 
instead of $-33$ per cent.  Such a model would of course be unrealistically pessimistic for surveys covering a broad range in galaxy luminosity, since 
by taking the $2\sigma$ upper limit in each luminosity bin and averaging these numbers we obtain a $>2\sigma$ upper limit on the average.

\subsection{Constraining $\sigma_8$}

Finally we come to the issue of greatest practical importance in the near term: how should one account for GI contamination in $\sigma_8$ measurements 
from cosmic shear?  We have already seen that if our more pessimistic models are correct then the contamination may be significant compared to the 
uncertainties in some of the recent measurements.  In this case it is essential to correct for the GI effect and include the range of allowed GI models 
in determining the error bars on $\sigma_8$.  Since the present surveys are measuring a single amplitude $\sigma_8$ (with some dependence on other 
cosmological parameters such as $\Omega_m$) what we really need is a probability distribution for $\Delta\sigma_8$ (which also may depend on other 
parameters) over which one can marginalize.  In future surveys that measure several parameters we will need a multivariate distribution.  Therefore we 
view the model here as a first step: it leaves much room for future work to reduce the uncertainties in $\Delta\sigma_8$, particularly to pin down the 
several-$\sigma$ tails of the distribution (which are treated rather crudely here), and to extend the model to incorporate multiple parameters.

In constructing this distribution for $\Delta\sigma_8$, we first note that the error bar on the GI contamination is highly asymmetric due to the nature 
of the redshift extrapolation: if the redshift exponent $\gamma$ is very negative or zero there is very little contamination, but if it is positive 
then it is possible to have very large contamination.  In all cases the central model predicts relatively small contamination (a few per cent) but the 
pessimistic model based on the $2\sigma$ upper limit to $\gamma$ could be a factor of several worse, and the optimistic models are a factor of several 
better.  Therefore we recommend that for measurements of $\sigma_8$ one should marginalize over contamination $\Delta\sigma_8$ with a lognormal 
probability distribution.  To be more explicit, we recommend the following prescription: for the distribution of galaxy luminosities, redshifts, and 
colours in a particular cosmic shear survey, compute the pessimistic (A) and central (B) power spectra $C_l^{GI}({\rm A},{\rm B})$.  Then compute the 
induced error $\Delta\sigma_8({\rm A},{\rm B})$ for these two models.  One can then write
\beq
\sigma_8({\rm observed}) = \sigma_8({\rm true}) -x,
\label{eq:x-model}
\eeq
where $x$ is lognormally distributed:
\beq
P(x) = \frac 1{\sqrt{2\pi}\;\sigma x}\exp\left[-\frac 1{2\sigma^2}\left( \ln \frac x{x_0}\right)^2 \right]
\label{eq:pofx}
\eeq
for $x>0$ and $P(x)=0$ for $x<0$.  The median of this distribution is the central model, i.e.
$x_0 = -\Delta\sigma_8({\rm B})$,
and the standard deviation is chosen to place the pessimistic model at $2\sigma$:
\beq
\sigma = \frac 12 \ln \frac{\Delta\sigma_8({\rm A})}{\Delta\sigma_8({\rm B})}.
\eeq
Usually one uses a Markov chain Monte Carlo (MCMC) method to estimate cosmological parameters from a combination of data sets; in this case one should 
marginalize over $x$ with the prior given by Eq.~(\ref{eq:pofx}).  This could be done by including $x$ as a nuisance parameter in the MCMC, or 
(probably faster) by including the integral over $x$ as part of the cosmic shear likelihood function.

Note that this method has not explicitly used the optimistic models in constructing the probability distribution for $x$, assuming instead that the 
distribution in $\ln x$ is symmetric.  In practice this is probably not a serious deficiency for two reasons.  First, for the cases we have 
investigated, the ratio of contamination for model A to model B is indeed similar to that of model B to model D (5.1 vs. 4.3 for $R<24$, $l=500$; 4.1 
vs. 3.0 for $R<23$, $l=500$), that is the error in ``log contamination'' is close to symmetric.  Secondly, the error in $\sigma_8$ in the optimistic 
and very optimistic models is usually $<0.01$, which is negligible compared to the purely statistical errors from the current generation of weak 
lensing surveys.  This means that while the details of the pessimistic tail of the distribution matter (because they affect whether cosmic shear can 
rule out high $\sigma_8$), the details of the optimistic tail do not -- the optimistic tail might as well be ``piled up'' at zero contamination.  

We believe the method described here is adequate for the current generation of cosmic shear surveys in which the cosmological constraint is a single 
amplitude, generally reported as a value of $\sigma_8$ with some dependence on $\Omega_m$.  However, it will not be adequate for future surveys that 
will measure the redshift and scale dependence of the signal, in which the GI contamination must be described by more than one number.  Also these 
surveys will push the statistical errors on the cosmic shear signal to the $<1$ per cent level, i.e. according to the models presented here the GI 
contamination will dominate over statistical uncertainty.  In order to make use of this data, we will need more external information to better 
constrain the GI models, or use internal information from the cosmic shear surveys themselves to simultaneously constrain the pure lensing signal, the 
GI signal, and (if applicable) the II signal.  Strategies for this are discussed in the literature \citep{2004PhRvD..70f3526H, 2005A&A...441...47K, 
2007arXiv0705.0166B}.

In summary, we have constructed a basic model for the GI contamination that can be practically integrated into current and near-future lensing 
constraints on $\sigma_8$.  It is a minimal model, making the simplest assumptions in some cases (e.g. power law scale, luminosity, and redshift 
dependence).  Nevertheless, it fulfills the basic criteria of being consistent with the data and covering the range of allowed values of the most 
important uncertain parameter ($\gamma$).  Aside from improving the statistical uncertainties on GI model parameters, future work covering a range of 
redshifts would also improve the robustness of the model by enabling us to constrain more model parameters, such as deviations from the power law 
$\sim(1+z)^\gamma$ or different values of $\alpha$ and $\gamma$ for blue and red galaxies.

\section{Conclusions}\label{S:conclusions}

In this work, we have attempted to further characterize the correlation between the intrinsic shear and the density field on both small and large 
scales.  To this end, we have characterized the strength of this alignment as a function of galaxy transverse separation, luminosity (focusing on the 
bright end, where it is most prominent), colour, and redshift to $z \sim 0.6$.  In addition, we have established that the GI amplitude of LRGs is not 
significantly different for \BCG\ and \nonBCG\ galaxies for scales above the virial radius, though we were unable to verify the degree to which this 
statement is true for cases of large luminosity gap. We found that the shape-density alignment signal for red galaxies increases strongly as a function 
of the luminosity. The scale dependence $w_{g+}(r_p)\propto r^{-0.73\pm 0.17}$ is very similar to that of the matter power spectrum, which is expected 
if galaxy alignment is determined by the local tidal field. Our findings that the density-shape alignment effect is present for \nonBCG s up to large 
scales implies that our original hypothesis that the effect is due only to \BCG s needs to be revised.

There is no detection of GI correlation for blue galaxies, with the possible exception of the very brightest bin (L6 blue).  In this case there is a 
$\sim 2.4\sigma$ detection if one fixes the scaling with $r_p$ to the value for the red galaxies.  If one does a ``blind'' search over power laws 
$r_p^\alpha$ (as done in Section~\ref{ss:sdss-main}) the signal is not statistically significant.  When combined with the fact that we have no 
detection in the lower-luminosity blue bins which have many more galaxies, we believe that the L6 blue GI signal is not robust, and observationally 
many of these galaxies are quite close to the blue versus red division.  In any case the L6 blue galaxies are very rare (despite their luminosity, they 
comprise $<1$ per cent of the flux-limited SDSS Main sample!) and their measured GI signal would contaminate the toy cosmic shear survey considered in 
Section~\ref{S:contam} at the $<0.1$ per cent level if taken at face value.  Much more important for the cosmic shear programme are the numerous 
fainter ($L\sim L_\star$) blue galaxies.  The GI signal for these objects is consistent with zero, however if the signal is near our upper limits, they 
may affect the observed shear power spectrum by of order 10 per cent for surveys to $R=24$.  Reducing the uncertainties in the measurements for the 
$L\sim L_\star$ blue galaxies should be a priority for future work.  It would also be desirable to learn more about the GI correlation 
$w_{\delta+}(r_p)$ at very small scales where linear galaxy biasing (which underlies the methodology of Section~\ref{S:fits}) breaks down; this would 
likely involve both the $w_{g+}(r_p)$ measurements presented in Appendix~\ref{app:data} and simulation or halo model results on the relation between 
galaxies and mass at small scales.

We have presented several models of the GI signal in Section~\ref{S:contam}, spanning the range from pessimistic assumptions (model A) to 
optimistic (model D).  We believe these models will be useful for others in comparing against other datasets and/or estimating levels of 
contamination in various surveys.  In particular it would be useful
for lensing surveys to repeat the calculations in
Section~\ref{S:contam}, taking  
into account their actual redshift/luminosity/colour distribution
instead of the toy distributions used here. 

These results should also be useful in comparing against simulations to determine
the physical cause behind this effect, especially in conjunction with
previous results on II contamination and future work on contamination
of three-point functions, which may give 
additional power to discriminate between intrinsic alignment models.
A fuller understanding of the physics behind these effects may allow us
to extend these fitting relations to higher redshift than
is currently allowed by the data.

Methods have been proposed \citep[e.g., ][]{2005A&A...441...47K} to
remove intrinsic alignment 
contamination using parameterized GI and II correlation models.  Our
determination of the redshift evolution and luminosity scaling of GI
correlations will help make these methods more feasible in practice.

Finally, the GI models presented here (and any improved models that incorporate future observations) may be used to
determine criteria for excluding galaxies from future cosmic shear
surveys to obtain a galaxy sample with the lowest possible level of
intrinsic alignment contamination.  Surveys with imaging data in
multiple bands should fairly easily be able to remove the bright red
galaxies that seem to show the strongest GI contamination.  Future
work with simulations will be necessary to determine the efficacy of
this plan.  This plan is most likely to be effective if the blue
galaxy contamination is found in future work to be negligibly small;
if this is not the case, then more sophisticated methods to remove GI such as 
templates \citep{2005A&A...441...47K} and separation based on redshift dependence \citep{2004PhRvD..70f3526H} will be essential to realising the 
promise of the cosmic shear programmes.

\section*{Acknowledgments}

C.H. is a John Bahcall Fellow in Astrophysics at the Institute for Advanced Study. R.M. is supported by NASA through Hubble Fellowship grant 
\#HST-HF-01199.02-A awarded by the Space Telescope Science Institute, which is operated by the Association of Universities for Research in Astronomy, 
Inc., for NASA, under contract NAS 5-26555.  M.I. acknowledges partial support from the Hoblitzelle foundation and a Clark award at the University of 
Texas, Dallas.  U.S. is supported by the Packard Foundation, NASA NAG5-1993 and NSF CAREER-0132953.

We thank Christian Wolf for clarification regarding the templates
used by the COMBO-17 survey in \cite{2003A&A...401...73W}.  We would also like to thank the anonymous referee for helpful comments.

Funding for the creation and distribution of the SDSS Archive has been 
provided by the Alfred P. Sloan Foundation, the Participating 
Institutions, the National Aeronautics and Space Administration, the 
National Science Foundation, the U.S. Department of Energy, the Japanese 
Monbukagakusho, and the Max Planck Society. The SDSS Web site is 
{\tt http://www.sdss.org/}.

The SDSS is managed by the Astrophysical Research Consortium (ARC) for the 
Participating Institutions. The Participating Institutions are The 
University of Chicago, Fermilab, the Institute for Advanced Study, the 
Japan Participation Group, The Johns Hopkins University, the Korean 
Scientist Group, Los Alamos National Laboratory, the Max-Planck-Institute 
for Astronomy (MPIA), the Max-Planck-Institute for Astrophysics (MPA), New 
Mexico State University, University of Pittsburgh, University of 
Portsmouth, Princeton University, the United States Naval Observatory, and 
the University of Washington.

\appendix

\section{Galaxy density-shape correlation function data}
\label{app:data}

This appendix lists the correlation functions $w_{g+}(r_p)$ for the samples used in our fits, and their correlation coefficients.  The table headers 
describe the subsamples used to trace the intrinsic shear field; the galaxy density (``$g$'') was traced using the full sample (i.e. SDSS Main, SDSS 
LRG low-$z$, SDSS LRG high-$z$, or 2SLAQ).  It is the bias of the latter (described in Section~\ref{S:bias}) that should be used to convert to 
$w_{\delta+}(r_p)$.  Table~\ref{tab:data} lists the measured correlation functions and their error bars, while Table~\ref{tab:corr} shows the 
correlation coefficients $\rho[w_{\delta+}(r_p),w_{\delta+}(r'_p)]$ between different radial bins for the same sample.  The innermost bin from the 
2SLAQ-bright sample did not have enough data to determine a bootstrap error, so it and its correlation matrix elements are marked with a 
``*'' in these tables.

\begin{table*}
\caption{\label{tab:data}The correlation functions $w_{g+}(r_p)$ for each sample.  The first column shows the range of $r_p$, and the remaining 
columns show $w_{g+}(r_p)$ and its 1$\sigma$ uncertainty (i.e. square root of diagonal covariance matrix element).  Units are $h^{-1}\,$Mpc for all 
columns.}
\begin{tabular}{crrrrrr}
\hline\hline
$r_{p,\rm min}$--$r_{p,\rm max}$ & Main.L3.blue & Main.L4.blue & Main.L5.blue & Main.L6.blue & Main.L3.red & Main.L4.red \\
\hline
0.3--0.6 & $ 0.145\pm 0.198$ & $-0.177\pm 0.194$ & $ 0.494\pm 0.443$ & $-1.127\pm 6.011$ & $ 0.117\pm 0.330$ & $ 0.032\pm 0.212$ \\
0.6--1.2 & $-0.081\pm 0.129$ & $ 0.118\pm 0.108$ & $-0.106\pm 0.215$ & $ 1.604\pm 1.854$ & $ 0.259\pm 0.147$ & $ 0.070\pm 0.100$ \\
1.2--2.4 & $-0.038\pm 0.069$ & $-0.020\pm 0.067$ & $ 0.100\pm 0.134$ & $ 1.588\pm 0.761$ & $ 0.003\pm 0.101$ & $-0.005\pm 0.053$ \\
2.4--4.7 & $ 0.033\pm 0.035$ & $-0.012\pm 0.035$ & $ 0.018\pm 0.088$ & $ 0.590\pm 0.653$ & $ 0.053\pm 0.057$ & $ 0.008\pm 0.053$ \\
4.7--7.5 & $ 0.017\pm 0.034$ & $ 0.061\pm 0.036$ & $-0.030\pm 0.060$ & $ 0.948\pm 0.395$ & $ 0.103\pm 0.048$ & $ 0.024\pm 0.036$ \\
7.5--11.9 & $ 0.059\pm 0.042$ & $ 0.016\pm 0.030$ & $-0.040\pm 0.050$ & $ 0.585\pm 0.329$ & $ 0.110\pm 0.042$ & $ 0.036\pm 0.026$ \\
11.9--18.9 & $-0.000\pm 0.032$ & $ 0.036\pm 0.026$ & $-0.048\pm 0.046$ & $ 0.227\pm 0.313$ & $ 0.073\pm 0.041$ & $ 0.014\pm 0.024$ \\
18.9--29.9 & $-0.010\pm 0.025$ & $ 0.015\pm 0.021$ & $-0.000\pm 0.045$ & $ 0.177\pm 0.214$ & $ 0.059\pm 0.040$ & $ 0.011\pm 0.018$ \\
29.9--47.4 & $ 0.003\pm 0.020$ & $ 0.007\pm 0.021$ & $-0.007\pm 0.029$ & $ 0.176\pm 0.155$ & $ 0.010\pm 0.034$ & $ 0.008\pm 0.013$ \\
47.4--59.6 & $ 0.017\pm 0.026$ & $ 0.003\pm 0.019$ & $ 0.007\pm 0.034$ & $-0.046\pm 0.164$ & $-0.044\pm 0.035$ & $ 0.010\pm 0.019$ \\
\hline
$r_{p,\rm min}$--$r_{p,\rm max}$ & Main.L5.red & Main.L6.red & LRG1,low-$z$ & LRG2,low-$z$ & LRG3,low-$z$ & LRG1,high-$z$ \\
\hline
0.3--0.6 & $ 0.744\pm 0.224$ & $ 7.676\pm 1.387$ & $ 5.075\pm 9.556$ & $ 1.365\pm 8.740$ & $ 2.975\pm 6.866$ & $-1.277\pm 8.148$ \\
0.6--1.2 & $ 0.278\pm 0.127$ & $ 2.906\pm 0.644$ & $ 3.590\pm 1.753$ & $ 1.007\pm 2.188$ & $ 1.714\pm 3.181$ & $ 2.784\pm 1.583$ \\
1.2--2.4 & $ 0.212\pm 0.087$ & $ 0.956\pm 0.288$ & $ 1.978\pm 0.873$ & $-0.216\pm 0.606$ & $ 1.633\pm 0.769$ & $ 0.370\pm 0.526$ \\
2.4--4.7 & $ 0.149\pm 0.051$ & $ 0.619\pm 0.158$ & $ 1.033\pm 0.327$ & $ 0.105\pm 0.312$ & $ 0.957\pm 0.302$ & $ 0.693\pm 0.242$ \\
4.7--7.5 & $ 0.039\pm 0.036$ & $ 0.420\pm 0.132$ & $ 0.270\pm 0.189$ & $ 0.182\pm 0.187$ & $ 0.362\pm 0.190$ & $ 0.441\pm 0.191$ \\
7.5--11.9 & $ 0.062\pm 0.028$ & $ 0.243\pm 0.107$ & $ 0.274\pm 0.130$ & $ 0.078\pm 0.164$ & $ 0.708\pm 0.135$ & $ 0.388\pm 0.133$ \\
11.9--18.9 & $ 0.046\pm 0.030$ & $ 0.306\pm 0.076$ & $ 0.339\pm 0.100$ & $-0.001\pm 0.100$ & $ 0.250\pm 0.087$ & $ 0.349\pm 0.076$ \\
18.9--29.9 & $ 0.092\pm 0.027$ & $ 0.239\pm 0.063$ & $ 0.135\pm 0.063$ & $ 0.035\pm 0.059$ & $ 0.209\pm 0.062$ & $ 0.209\pm 0.071$ \\
29.9--47.4 & $ 0.036\pm 0.018$ & $ 0.120\pm 0.060$ & $ 0.053\pm 0.050$ & $ 0.106\pm 0.055$ & $ 0.164\pm 0.043$ & $ 0.138\pm 0.035$ \\
47.4--59.6 & $ 0.010\pm 0.016$ & $ 0.016\pm 0.038$ & $-0.036\pm 0.057$ & $ 0.115\pm 0.045$ & $ 0.092\pm 0.041$ & $ 0.107\pm 0.047$ \\
\hline
$r_{p,\rm min}$--$r_{p,\rm max}$ & LRG2,high-$z$ & LRG3,high-$z$ & 2SLAQ-faint & 2SLAQ-bright &  &  \\
\hline
0.3--0.6 & $-7.429\pm 7.872$ & $21.379\pm 9.708$ & $-3.713\pm 8.009$ & * &  &  \\
0.6--1.2 & $ 4.603\pm 3.609$ & $ 7.346\pm 3.589$ & $-3.235\pm 3.722$ & $ 5.380\pm 3.223$ &  &  \\
1.2--2.4 & $ 2.940\pm 1.125$ & $ 2.634\pm 0.792$ & $ 0.425\pm 1.327$ & $ 2.717\pm 1.287$ &  &  \\
2.4--4.7 & $ 1.950\pm 0.444$ & $ 1.150\pm 0.406$ & $-0.155\pm 0.499$ & $ 0.598\pm 0.621$ &  &  \\
4.7--7.5 & $ 0.629\pm 0.296$ & $ 0.890\pm 0.308$ & $ 0.506\pm 0.385$ & $-0.230\pm 0.396$ &  &  \\
7.5--11.9 & $ 0.518\pm 0.208$ & $ 0.539\pm 0.155$ & $-0.335\pm 0.333$ & $ 0.469\pm 0.214$ &  &  \\
11.9--18.9 & $ 0.592\pm 0.113$ & $ 0.365\pm 0.107$ & $-0.168\pm 0.161$ & $ 0.087\pm 0.155$ &  &  \\
18.9--29.9 & $ 0.344\pm 0.103$ & $ 0.290\pm 0.084$ & $ 0.004\pm 0.172$ & $-0.020\pm 0.138$ &  &  \\
29.9--47.4 & $ 0.094\pm 0.080$ & $ 0.257\pm 0.054$ & $-0.085\pm 0.157$ & $ 0.209\pm 0.116$ &  &  \\
47.4--59.6 & $ 0.103\pm 0.077$ & $ 0.144\pm 0.060$ & $-0.129\pm 0.207$ & $ 0.112\pm 0.143$ &  &  \\
\hline\hline
\end{tabular}
\end{table*}

\begin{table*}
\caption{\label{tab:corr}Correlation matrices for the data shown in Table~\ref{tab:data}.  The units are per cents (i.e. all correlation matrix 
elements have been multiplied by 100).  In each of the 4 blocks of this table, there is an upper (above-diagonal) triangle, which applies to the
sample designated ``upper,'' and a lower (below-diagonal) triangle, which applies to the sample designated ``lower.''  Note that the diagonal 
correlation coefficients are equal to unity (100 per cent).}
\begin{tabular}{rrrrrrrrrrrrrrrrrrrrr}
\hline\hline
\multicolumn{10}{c}{\mbox{Main.L3.blue (upper); Main.L4.blue (lower)}}
  & & \multicolumn{10}{c}{\mbox{Main.L5.blue (upper); Main.L6.blue (lower)}}  \\
\hline
$100$ & $\!\!$$4$ & $\!\!$$9$ & $\!\!$$-10$ & $\!\!$$22$ & $\!\!$$41$ & $\!\!$$11$ & $\!\!$$47$ & $\!\!$$-1$ & $\!\!$$-15$ & &
$100$ & $\!\!$$18$ & $\!\!$$18$ & $\!\!$$6$ & $\!\!$$26$ & $\!\!$$7$ & $\!\!$$-9$ & $\!\!$$-10$ & $\!\!$$-11$ & $\!\!$$-12$ \\
$7$ & $\!\!$$100$ & $\!\!$$2$ & $\!\!$$-15$ & $\!\!$$-6$ & $\!\!$$16$ & $\!\!$$15$ & $\!\!$$1$ & $\!\!$$-12$ & $\!\!$$-19$ & &
$26$ & $\!\!$$100$ & $\!\!$$20$ & $\!\!$$-2$ & $\!\!$$10$ & $\!\!$$11$ & $\!\!$$1$ & $\!\!$$5$ & $\!\!$$1$ & $\!\!$$4$ \\
$6$ & $\!\!$$21$ & $\!\!$$100$ & $\!\!$$24$ & $\!\!$$-7$ & $\!\!$$-13$ & $\!\!$$0$ & $\!\!$$-2$ & $\!\!$$-33$ & $\!\!$$-12$ & &
$-14$ & $\!\!$$11$ & $\!\!$$100$ & $\!\!$$30$ & $\!\!$$25$ & $\!\!$$8$ & $\!\!$$8$ & $\!\!$$-2$ & $\!\!$$-10$ & $\!\!$$16$ \\
$-24$ & $\!\!$$12$ & $\!\!$$27$ & $\!\!$$100$ & $\!\!$$10$ & $\!\!$$3$ & $\!\!$$7$ & $\!\!$$8$ & $\!\!$$-7$ & $\!\!$$-16$ & &
$27$ & $\!\!$$37$ & $\!\!$$9$ & $\!\!$$100$ & $\!\!$$45$ & $\!\!$$40$ & $\!\!$$7$ & $\!\!$$6$ & $\!\!$$5$ & $\!\!$$0$ \\
$-18$ & $\!\!$$20$ & $\!\!$$22$ & $\!\!$$33$ & $\!\!$$100$ & $\!\!$$50$ & $\!\!$$24$ & $\!\!$$17$ & $\!\!$$3$ & $\!\!$$5$ & &
$40$ & $\!\!$$42$ & $\!\!$$15$ & $\!\!$$41$ & $\!\!$$100$ & $\!\!$$38$ & $\!\!$$-14$ & $\!\!$$-22$ & $\!\!$$-10$ & $\!\!$$6$ \\
$-13$ & $\!\!$$-10$ & $\!\!$$14$ & $\!\!$$-5$ & $\!\!$$17$ & $\!\!$$100$ & $\!\!$$56$ & $\!\!$$25$ & $\!\!$$1$ & $\!\!$$-8$ & &
$33$ & $\!\!$$23$ & $\!\!$$-16$ & $\!\!$$30$ & $\!\!$$46$ & $\!\!$$100$ & $\!\!$$35$ & $\!\!$$8$ & $\!\!$$-7$ & $\!\!$$-2$ \\
$2$ & $\!\!$$-3$ & $\!\!$$20$ & $\!\!$$3$ & $\!\!$$16$ & $\!\!$$26$ & $\!\!$$100$ & $\!\!$$26$ & $\!\!$$-3$ & $\!\!$$-22$ & &
$37$ & $\!\!$$15$ & $\!\!$$8$ & $\!\!$$12$ & $\!\!$$43$ & $\!\!$$37$ & $\!\!$$100$ & $\!\!$$65$ & $\!\!$$19$ & $\!\!$$2$ \\
$12$ & $\!\!$$-18$ & $\!\!$$12$ & $\!\!$$-16$ & $\!\!$$7$ & $\!\!$$11$ & $\!\!$$37$ & $\!\!$$100$ & $\!\!$$25$ & $\!\!$$-30$ & &
$35$ & $\!\!$$16$ & $\!\!$$5$ & $\!\!$$8$ & $\!\!$$15$ & $\!\!$$7$ & $\!\!$$56$ & $\!\!$$100$ & $\!\!$$29$ & $\!\!$$4$ \\
$1$ & $\!\!$$14$ & $\!\!$$8$ & $\!\!$$-7$ & $\!\!$$16$ & $\!\!$$8$ & $\!\!$$0$ & $\!\!$$20$ & $\!\!$$100$ & $\!\!$$51$ & &
$43$ & $\!\!$$13$ & $\!\!$$4$ & $\!\!$$0$ & $\!\!$$25$ & $\!\!$$9$ & $\!\!$$18$ & $\!\!$$29$ & $\!\!$$100$ & $\!\!$$47$ \\
$-11$ & $\!\!$$10$ & $\!\!$$-13$ & $\!\!$$20$ & $\!\!$$16$ & $\!\!$$-12$ & $\!\!$$7$ & $\!\!$$1$ & $\!\!$$49$ & $\!\!$$100$ & &
$6$ & $\!\!$$14$ & $\!\!$$-7$ & $\!\!$$5$ & $\!\!$$21$ & $\!\!$$-2$ & $\!\!$$4$ & $\!\!$$24$ & $\!\!$$39$ & $\!\!$$100$ \\
\hline
\multicolumn{10}{c}{\mbox{Main.L3.red (upper); Main.L4.red (lower)}}
  & & \multicolumn{10}{c}{\mbox{Main.L5.red (upper); Main.L6.red (lower)}}  \\
\hline
$100$ & $\!\!$$34$ & $\!\!$$-23$ & $\!\!$$-3$ & $\!\!$$6$ & $\!\!$$-22$ & $\!\!$$2$ & $\!\!$$-1$ & $\!\!$$19$ & $\!\!$$9$ & &
$100$ & $\!\!$$39$ & $\!\!$$0$ & $\!\!$$14$ & $\!\!$$13$ & $\!\!$$7$ & $\!\!$$-15$ & $\!\!$$-18$ & $\!\!$$-11$ & $\!\!$$8$ \\
$20$ & $\!\!$$100$ & $\!\!$$16$ & $\!\!$$8$ & $\!\!$$24$ & $\!\!$$-1$ & $\!\!$$-1$ & $\!\!$$-5$ & $\!\!$$3$ & $\!\!$$-9$ & &
$45$ & $\!\!$$100$ & $\!\!$$15$ & $\!\!$$22$ & $\!\!$$14$ & $\!\!$$13$ & $\!\!$$1$ & $\!\!$$9$ & $\!\!$$-7$ & $\!\!$$-24$ \\
$12$ & $\!\!$$8$ & $\!\!$$100$ & $\!\!$$8$ & $\!\!$$2$ & $\!\!$$26$ & $\!\!$$-5$ & $\!\!$$-5$ & $\!\!$$-4$ & $\!\!$$-7$ & &
$34$ & $\!\!$$47$ & $\!\!$$100$ & $\!\!$$37$ & $\!\!$$-18$ & $\!\!$$10$ & $\!\!$$-5$ & $\!\!$$-2$ & $\!\!$$18$ & $\!\!$$-17$ \\
$13$ & $\!\!$$-19$ & $\!\!$$44$ & $\!\!$$100$ & $\!\!$$15$ & $\!\!$$7$ & $\!\!$$18$ & $\!\!$$5$ & $\!\!$$-8$ & $\!\!$$-6$ & &
$40$ & $\!\!$$14$ & $\!\!$$37$ & $\!\!$$100$ & $\!\!$$-21$ & $\!\!$$-14$ & $\!\!$$-14$ & $\!\!$$-20$ & $\!\!$$2$ & $\!\!$$-12$ \\
$-22$ & $\!\!$$-29$ & $\!\!$$11$ & $\!\!$$55$ & $\!\!$$100$ & $\!\!$$27$ & $\!\!$$7$ & $\!\!$$6$ & $\!\!$$-6$ & $\!\!$$-18$ & &
$33$ & $\!\!$$31$ & $\!\!$$13$ & $\!\!$$24$ & $\!\!$$100$ & $\!\!$$36$ & $\!\!$$0$ & $\!\!$$16$ & $\!\!$$-2$ & $\!\!$$1$ \\
$-7$ & $\!\!$$0$ & $\!\!$$8$ & $\!\!$$16$ & $\!\!$$46$ & $\!\!$$100$ & $\!\!$$50$ & $\!\!$$28$ & $\!\!$$4$ & $\!\!$$1$ & &
$12$ & $\!\!$$-16$ & $\!\!$$11$ & $\!\!$$13$ & $\!\!$$39$ & $\!\!$$100$ & $\!\!$$43$ & $\!\!$$22$ & $\!\!$$15$ & $\!\!$$-25$ \\
$-6$ & $\!\!$$12$ & $\!\!$$1$ & $\!\!$$8$ & $\!\!$$4$ & $\!\!$$19$ & $\!\!$$100$ & $\!\!$$69$ & $\!\!$$8$ & $\!\!$$-7$ & &
$14$ & $\!\!$$11$ & $\!\!$$8$ & $\!\!$$0$ & $\!\!$$29$ & $\!\!$$53$ & $\!\!$$100$ & $\!\!$$53$ & $\!\!$$32$ & $\!\!$$5$ \\
$0$ & $\!\!$$22$ & $\!\!$$-8$ & $\!\!$$1$ & $\!\!$$-2$ & $\!\!$$5$ & $\!\!$$63$ & $\!\!$$100$ & $\!\!$$24$ & $\!\!$$4$ & &
$12$ & $\!\!$$11$ & $\!\!$$17$ & $\!\!$$-9$ & $\!\!$$7$ & $\!\!$$17$ & $\!\!$$54$ & $\!\!$$100$ & $\!\!$$44$ & $\!\!$$0$ \\
$25$ & $\!\!$$24$ & $\!\!$$-3$ & $\!\!$$-2$ & $\!\!$$-15$ & $\!\!$$-6$ & $\!\!$$27$ & $\!\!$$46$ & $\!\!$$100$ & $\!\!$$40$ & &
$-5$ & $\!\!$$-2$ & $\!\!$$0$ & $\!\!$$-17$ & $\!\!$$-5$ & $\!\!$$8$ & $\!\!$$33$ & $\!\!$$51$ & $\!\!$$100$ & $\!\!$$20$ \\
$17$ & $\!\!$$-4$ & $\!\!$$23$ & $\!\!$$-1$ & $\!\!$$-11$ & $\!\!$$-13$ & $\!\!$$0$ & $\!\!$$-19$ & $\!\!$$18$ & $\!\!$$100$ & &
$11$ & $\!\!$$0$ & $\!\!$$17$ & $\!\!$$-5$ & $\!\!$$10$ & $\!\!$$7$ & $\!\!$$20$ & $\!\!$$20$ & $\!\!$$32$ & $\!\!$$100$ \\
\hline
\multicolumn{10}{c}{\mbox{LRG1,low-$z$ (upper); LRG2,low-$z$ (lower)}}
  & & \multicolumn{10}{c}{\mbox{LRG3,low-$z$ (upper); LRG1,high-$z$ (lower)}}  \\
\hline
$100$ & $\!\!$$-8$ & $\!\!$$-23$ & $\!\!$$-8$ & $\!\!$$11$ & $\!\!$$3$ & $\!\!$$-9$ & $\!\!$$-5$ & $\!\!$$-8$ & $\!\!$$9$ & &
$100$ & $\!\!$$-32$ & $\!\!$$5$ & $\!\!$$25$ & $\!\!$$-17$ & $\!\!$$4$ & $\!\!$$3$ & $\!\!$$-4$ & $\!\!$$-24$ & $\!\!$$-4$ \\
$-13$ & $\!\!$$100$ & $\!\!$$-6$ & $\!\!$$-6$ & $\!\!$$-19$ & $\!\!$$-12$ & $\!\!$$8$ & $\!\!$$9$ & $\!\!$$2$ & $\!\!$$7$ & &
$4$ & $\!\!$$100$ & $\!\!$$14$ & $\!\!$$-5$ & $\!\!$$4$ & $\!\!$$-11$ & $\!\!$$17$ & $\!\!$$2$ & $\!\!$$-15$ & $\!\!$$-25$ \\
$-2$ & $\!\!$$16$ & $\!\!$$100$ & $\!\!$$-9$ & $\!\!$$-14$ & $\!\!$$-8$ & $\!\!$$22$ & $\!\!$$9$ & $\!\!$$2$ & $\!\!$$-3$ & &
$7$ & $\!\!$$4$ & $\!\!$$100$ & $\!\!$$20$ & $\!\!$$-33$ & $\!\!$$-11$ & $\!\!$$-10$ & $\!\!$$3$ & $\!\!$$-21$ & $\!\!$$0$ \\
$-16$ & $\!\!$$24$ & $\!\!$$4$ & $\!\!$$100$ & $\!\!$$1$ & $\!\!$$23$ & $\!\!$$27$ & $\!\!$$12$ & $\!\!$$13$ & $\!\!$$19$ & &
$8$ & $\!\!$$3$ & $\!\!$$-13$ & $\!\!$$100$ & $\!\!$$9$ & $\!\!$$-33$ & $\!\!$$-8$ & $\!\!$$2$ & $\!\!$$-15$ & $\!\!$$-20$ \\
$-8$ & $\!\!$$-2$ & $\!\!$$23$ & $\!\!$$11$ & $\!\!$$100$ & $\!\!$$27$ & $\!\!$$13$ & $\!\!$$-6$ & $\!\!$$1$ & $\!\!$$-9$ & &
$1$ & $\!\!$$2$ & $\!\!$$-8$ & $\!\!$$11$ & $\!\!$$100$ & $\!\!$$13$ & $\!\!$$1$ & $\!\!$$2$ & $\!\!$$20$ & $\!\!$$4$ \\
$27$ & $\!\!$$-3$ & $\!\!$$-8$ & $\!\!$$-21$ & $\!\!$$4$ & $\!\!$$100$ & $\!\!$$11$ & $\!\!$$-6$ & $\!\!$$3$ & $\!\!$$-2$ & &
$-1$ & $\!\!$$11$ & $\!\!$$26$ & $\!\!$$-6$ & $\!\!$$21$ & $\!\!$$100$ & $\!\!$$-11$ & $\!\!$$-2$ & $\!\!$$26$ & $\!\!$$3$ \\
$27$ & $\!\!$$-14$ & $\!\!$$-12$ & $\!\!$$-10$ & $\!\!$$-12$ & $\!\!$$23$ & $\!\!$$100$ & $\!\!$$9$ & $\!\!$$-13$ & $\!\!$$1$ & &
$7$ & $\!\!$$30$ & $\!\!$$1$ & $\!\!$$16$ & $\!\!$$-15$ & $\!\!$$-11$ & $\!\!$$100$ & $\!\!$$3$ & $\!\!$$-20$ & $\!\!$$-11$ \\
$29$ & $\!\!$$-5$ & $\!\!$$9$ & $\!\!$$11$ & $\!\!$$6$ & $\!\!$$18$ & $\!\!$$26$ & $\!\!$$100$ & $\!\!$$35$ & $\!\!$$12$ & &
$-17$ & $\!\!$$8$ & $\!\!$$9$ & $\!\!$$20$ & $\!\!$$0$ & $\!\!$$19$ & $\!\!$$30$ & $\!\!$$100$ & $\!\!$$-11$ & $\!\!$$-20$ \\
$-9$ & $\!\!$$0$ & $\!\!$$5$ & $\!\!$$8$ & $\!\!$$7$ & $\!\!$$-12$ & $\!\!$$-3$ & $\!\!$$-10$ & $\!\!$$100$ & $\!\!$$33$ & &
$-27$ & $\!\!$$11$ & $\!\!$$3$ & $\!\!$$26$ & $\!\!$$-5$ & $\!\!$$2$ & $\!\!$$-5$ & $\!\!$$0$ & $\!\!$$100$ & $\!\!$$39$ \\
$8$ & $\!\!$$2$ & $\!\!$$15$ & $\!\!$$12$ & $\!\!$$0$ & $\!\!$$2$ & $\!\!$$12$ & $\!\!$$11$ & $\!\!$$9$ & $\!\!$$100$ & &
$5$ & $\!\!$$10$ & $\!\!$$-2$ & $\!\!$$2$ & $\!\!$$-15$ & $\!\!$$4$ & $\!\!$$1$ & $\!\!$$9$ & $\!\!$$42$ & $\!\!$$100$ \\
\hline
\multicolumn{10}{c}{\mbox{LRG2,high-$z$ (upper); LRG3,high-$z$ (lower)}}
  & & \multicolumn{10}{c}{\mbox{2SLAQ-faint (upper); 2SLAQ-bright (lower)}}  \\
\hline
$100$ & $\!\!$$10$ & $\!\!$$-3$ & $\!\!$$-10$ & $\!\!$$-5$ & $\!\!$$15$ & $\!\!$$-1$ & $\!\!$$-20$ & $\!\!$$-11$ & $\!\!$$6$ & &
$100$ & $\!\!$$-39$ & $\!\!$$-9$ & $\!\!$$-3$ & $\!\!$$13$ & $\!\!$$-30$ & $\!\!$$-16$ & $\!\!$$6$ & $\!\!$$-9$ & $\!\!$$0$ \\
$18$ & $\!\!$$100$ & $\!\!$$-20$ & $\!\!$$-16$ & $\!\!$$-22$ & $\!\!$$0$ & $\!\!$$-15$ & $\!\!$$-4$ & $\!\!$$2$ & $\!\!$$5$ & &
* & $\!\!$$100$ & $\!\!$$-7$ & $\!\!$$7$ & $\!\!$$-11$ & $\!\!$$28$ & $\!\!$$8$ & $\!\!$$-13$ & $\!\!$$5$ & $\!\!$$-15$ \\
$10$ & $\!\!$$2$ & $\!\!$$100$ & $\!\!$$22$ & $\!\!$$5$ & $\!\!$$-6$ & $\!\!$$10$ & $\!\!$$28$ & $\!\!$$16$ & $\!\!$$13$ & &
* & $\!\!$$13$ & $\!\!$$100$ & $\!\!$$27$ & $\!\!$$18$ & $\!\!$$-1$ & $\!\!$$-11$ & $\!\!$$13$ & $\!\!$$25$ & $\!\!$$27$ \\
$-6$ & $\!\!$$-26$ & $\!\!$$-1$ & $\!\!$$100$ & $\!\!$$-3$ & $\!\!$$14$ & $\!\!$$23$ & $\!\!$$9$ & $\!\!$$-5$ & $\!\!$$-11$ & &
* & $\!\!$$0$ & $\!\!$$-11$ & $\!\!$$100$ & $\!\!$$19$ & $\!\!$$2$ & $\!\!$$27$ & $\!\!$$6$ & $\!\!$$42$ & $\!\!$$18$ \\
$-18$ & $\!\!$$9$ & $\!\!$$-16$ & $\!\!$$9$ & $\!\!$$100$ & $\!\!$$18$ & $\!\!$$28$ & $\!\!$$-1$ & $\!\!$$22$ & $\!\!$$11$ & &
* & $\!\!$$4$ & $\!\!$$16$ & $\!\!$$30$ & $\!\!$$100$ & $\!\!$$25$ & $\!\!$$20$ & $\!\!$$35$ & $\!\!$$37$ & $\!\!$$32$ \\
$-11$ & $\!\!$$2$ & $\!\!$$14$ & $\!\!$$3$ & $\!\!$$20$ & $\!\!$$100$ & $\!\!$$41$ & $\!\!$$-9$ & $\!\!$$-1$ & $\!\!$$-8$ & &
* & $\!\!$$5$ & $\!\!$$6$ & $\!\!$$-18$ & $\!\!$$9$ & $\!\!$$100$ & $\!\!$$27$ & $\!\!$$19$ & $\!\!$$7$ & $\!\!$$-8$ \\
$-9$ & $\!\!$$-19$ & $\!\!$$5$ & $\!\!$$11$ & $\!\!$$7$ & $\!\!$$28$ & $\!\!$$100$ & $\!\!$$0$ & $\!\!$$1$ & $\!\!$$11$ & &
* & $\!\!$$3$ & $\!\!$$13$ & $\!\!$$31$ & $\!\!$$5$ & $\!\!$$-2$ & $\!\!$$100$ & $\!\!$$44$ & $\!\!$$26$ & $\!\!$$9$ \\
$3$ & $\!\!$$5$ & $\!\!$$0$ & $\!\!$$10$ & $\!\!$$27$ & $\!\!$$12$ & $\!\!$$1$ & $\!\!$$100$ & $\!\!$$53$ & $\!\!$$44$ & &
* & $\!\!$$0$ & $\!\!$$-23$ & $\!\!$$13$ & $\!\!$$-8$ & $\!\!$$-17$ & $\!\!$$24$ & $\!\!$$100$ & $\!\!$$54$ & $\!\!$$37$ \\
$3$ & $\!\!$$17$ & $\!\!$$23$ & $\!\!$$14$ & $\!\!$$4$ & $\!\!$$11$ & $\!\!$$15$ & $\!\!$$4$ & $\!\!$$100$ & $\!\!$$33$ & &
* & $\!\!$$-17$ & $\!\!$$3$ & $\!\!$$13$ & $\!\!$$1$ & $\!\!$$-30$ & $\!\!$$20$ & $\!\!$$58$ & $\!\!$$100$ & $\!\!$$59$ \\
$2$ & $\!\!$$29$ & $\!\!$$6$ & $\!\!$$31$ & $\!\!$$21$ & $\!\!$$-17$ & $\!\!$$-6$ & $\!\!$$16$ & $\!\!$$16$ & $\!\!$$100$ & &
* & $\!\!$$1$ & $\!\!$$-32$ & $\!\!$$0$ & $\!\!$$18$ & $\!\!$$31$ & $\!\!$$-3$ & $\!\!$$-5$ & $\!\!$$-1$ & $\!\!$$100$ \\
\hline\hline
\end{tabular}
\end{table*}

\section{Relation of lensing shear-intrinsic shear and density-intrinsic shear correlation functions}
\label{app:heymans}

The purpose of this appendix is to relate the lensing shear-intrinsic shear correlation function $\langle \bgamma^G(z_s)\cdot 
\bgamma^I(z_1)\rangle_\theta$ 
measured in simulations by \citet{2006MNRAS.371..750H} to the density-intrinsic shear correlation function $w_{\delta+}(r_p)$ considered in this paper.  
In particular we would like to know whether the \citet{2006MNRAS.371..750H} are consistent with our observations.  In this appendix, we will use $z_s$ 
for the source-plane redshift and $z_1$ for the lens-plane redshift (i.e. the redshift at which GI contamination is being assessed) for consistency 
with \citet{2006MNRAS.371..750H}; however for consistency with our paper we will write $\bgamma^I$ where \citet{2006MNRAS.371..750H} would write $e$ 
[defined by their Eq.~(4); note that the ellipticity components before correction for the shear responsivity, which we denote $e_i$, are denoted by 
$\epsilon_i$ in \citet{2006MNRAS.371..750H}].

Our conversion proceeds in two steps: first from $\langle \bgamma^G(z_s)\cdot\bgamma^I(z_1)\rangle_\theta$ to
$\langle \kappa(z_s) \gamma^I_+(z_1)\rangle_\theta$, and then from $\langle \kappa(z_s) 
\gamma^I_+(z_1)\rangle_\theta$ to $w_{\delta+}(r_p)$.  The first step is a straightforward but tedious calculus exercise, since $\bgamma^G$ and 
$\kappa$ 
are different derivatives of the lensing potential and hence are simply related in Fourier space; the second step is trivial and involves the usual 
convergence to surface density conversion from galaxy-galaxy lensing studies.

For our first step, we recall that the shear components $+,\times$ can be easily converted to $E$ and $B$-modes in Fourier space by
\beq
\gamma^G_+({\bmath l})\pm\rmi\gamma^G_\times({\bmath l}) = \rme^{2\rmi\phi({\bmath l})}[\gamma^G_E({\bmath l})+\rmi 
\gamma^G_B({\bmath l})],
\label{eq:eb}
\eeq
where $\phi({\bmath l})=\arctan(l_y/l_x)$ is the position angle of ${\bmath l}$, and a similar formula is written for the intrinsic shear.  The 
convergence-density correlation function is then the Fourier transform of the cross-power spectrum,
\beqa
\langle\kappa(\btheta)\gamma^I_+({\bmath 0})\rangle &=& \int \frac{\rmd^2{\bmath l}}{(2\pi)^2} C_l^{\kappa\bgamma^I} \rme^{i{\bmath l}\cdot\btheta}
\rme^{2\rmi\phi({\bmath l})} \nonumber \\
&=& -\int_0^\infty \frac{l\,\rmd l}{2\pi} C_l^{\kappa\bgamma^I} J_2(l\theta).
\label{eq:he2}
\eeqa
Here we have assumed parity invariance, i.e. that the convergence is correlated only with the $E$-mode of the intrinsic shear.  (The latter may have a 
$B$-mode but this cannot be correlated with a scalar.)  We have also chosen $\btheta$ to lie along the $x$-axis (i.e. the axis along which stretching 
corresponds to positive $+$-component shear) and used the integral representation of the Bessel function (Eq.~8.411 of \citealt{1994tisp.book.....G}).  
A similar result holds for the $\gamma^G\gamma^I$ correlation:
\beq
\langle\bgamma^G(\btheta)\cdot\bgamma^I({\bmath 0})\rangle = \int_0^\infty \frac{l\,\rmd l}{2\pi} C_l^{\kappa\bgamma^I} J_0(l\theta);
\label{eq:he0}
\eeq
once again we have used only the $E$-mode since the lensing shear $\bgamma^G$ is derived from the scalar density field and has only $E$-modes (aside 
from small corrections due to non-weak shear or multiple deflections).  We have replaced $C_l^{\bgamma^G\bgamma^I}\rightarrow C_l^{\kappa\bgamma^I}$ 
since $\bgamma^G_E$ and $\kappa$ have equal numerical values mode by mode (they are both second derivatives of the lensing potential), and we have a 
$J_0$ function instead of $-J_2$ because there is no factor of $\rme^{2\rmi\phi({\bmath l})}$ in this equation [the dot product of the lensing and 
intrinsic shears is unaffected by the angle-$\phi({\bmath l})$ rotation of Eq.~(\ref{eq:eb})].

The conversion between Eq.~(\ref{eq:he2}) and Eq.~(\ref{eq:he0}) proceeds as follows.  The cross-power spectrum $C_l^{\kappa\bgamma^I}$ can be obtained 
from Eq.~(\ref{eq:he0}) by the Hankel transform pair,
\beq
C_l^{\kappa\bgamma^I}  = 2\pi \int_0^\infty \langle\bgamma^G(\bvartheta)\cdot\gamma^I({\bmath 0})\rangle J_0(l\vartheta) \vartheta\,\rmd\vartheta,
\eeq
from which we find
\beq
\langle\kappa(\btheta)\gamma^I_+({\bmath 0})\rangle = -\int_0^\infty
\langle\bgamma^G(\bvartheta)\cdot\bgamma^I({\bmath 0})\rangle
{\cal G}(\theta,\vartheta) \vartheta\, \rmd\vartheta,
\eeq
where the Green's function is
\beq
{\cal G}(\theta,\vartheta) \equiv \int_0^\infty \rmd l\,lJ_0(l\vartheta)J_2(l\theta).
\eeq
This Green's function has a closed-form analytical expression that can be obtained using the recursion, differentiation, and orthonormality relations 
for Bessel functions:
\beqa
{\cal G}(\theta,\vartheta) &=& \int_0^\infty \rmd l\,lJ_0(l\vartheta)\left[-J_0(l\theta)+\frac 2{l\theta}J_1(l\theta)\right]
\nonumber \\
&=& -\int_0^\infty \rmd l\,lJ_0(l\vartheta)J_0(l\theta) + \frac 2\theta\int_0^\infty \rmd l\,J_0(l\vartheta)J_1(l\theta)
\nonumber \\
&=& -\frac{\delta(\theta-\vartheta)}\theta + \frac 2\theta\int_0^\infty \rmd l\,\int_\vartheta^\infty \rmd w\;lJ_1(lw) J_1(l\theta)
\nonumber \\
&=& -\frac{\delta(\theta-\vartheta)}\theta + \frac 2\theta \int_\vartheta^\infty \rmd w \frac{\delta(w-\theta)}\theta
\nonumber \\
&=& -\frac{\delta(\theta-\vartheta)}\theta + \frac 2{\theta^2} \Theta(\theta-\vartheta),
\eeqa
where $\Theta$ is the Heaviside step function.  (The third line has used the fact that $-J_1$ is the derivative of $J_0$.)  Therefore
\beqa
\langle\kappa(\btheta)\gamma^I_+({\bmath 0})\rangle &=& -\langle\bgamma^G(\theta)\cdot\bgamma^I({\bmath 0})\rangle
\nonumber \\ &&
+ \frac 2{\theta^2}\int_0^\theta \langle\bgamma^G(\bvartheta)\cdot\bgamma^I({\bmath 0})\rangle\vartheta\, \rmd\vartheta.
\label{eq:conv1}
\eeqa
\citet{2006MNRAS.371..750H} fit their results to the functional form
\beq
\langle\bgamma^G(z_s)\cdot\bgamma^I(z_1)\rangle_\theta = \frac{A{\cal E}}{\theta+\theta_0},
\eeq
where $A$ and $\theta_0$ are free parameters and ${\cal E}=D_lD_{ls}/D_s$ is the lensing strength.
Plugging this into Eq.~(\ref{eq:conv1}) yields
\beq
\langle\kappa(z_s)\gamma^I_+(z_1)\rangle_\theta = A{\cal E}\left( \frac 1{\theta+\theta_0}
-\frac 2\theta + \frac{2\theta_0}{\theta^2}\ln\frac{\theta+\theta_0}{\theta_0}\right).
\label{eq:conv2}
\eeq

Our second step is to convert this into a density-intrinsic shear correlation function.  To do this, we recall the Born (single-deflection) 
approximation
for the convergence,
\beq
\kappa(z_s) = \frac{3\Omega_mH_0^2}{2c^2} \int \rmd r\, {\cal E} \delta(r),
\eeq
where $\delta$ is the density perturbation.  If the correlations between $\kappa(z_s)$ and the intrinsic shear field at a particular redshift $z_l$ 
arise near $z_l$, then we may change variables from distance from the observer $r$ to radial separation $\Pi=r-r(z_l)$, multiply both sides by the 
intrinsic shear field, and take the average:
\beq
\langle\kappa(z_s)\gamma^I_+(z_1)\rangle_\theta = \frac{3\Omega_mH_0^2}{2c^2}{\cal E}\int \rmd\Pi\,\xi_{\delta+}(r_p,\Pi),
\eeq
where $r_p = r\theta$ is the separation and we can pull ${\cal E}$ (which varies slowly with lens redshift) out of the integral because only a small 
range at $z\approx z_l$ contributes.  The last integral is $w_{\delta+}(r_p)$, so combining with Eq.~(\ref{eq:conv2}) we have
\beq
w_{\delta+}(r_p) = \frac{2c^2A}{3\Omega_mH_0^2}
\left( \frac 1{\theta+\theta_0}
-\frac 2\theta + \frac{2\theta_0}{\theta^2}\ln\frac{\theta+\theta_0}{\theta_0}\right).
\label{eq:conv3}
\eeq
This is the equation we have used to compare the \citet{2006MNRAS.371..750H} results to ours in Section~\ref{ss:comparison}.

\end{document}